\documentclass[useAMS,usenatbib]{mn2e}
\usepackage{graphicx}
\usepackage{amssymb}
\usepackage{amsmath}

\def\jref@jnl#1{{\rm#1}}

\def\aj{\jref@jnl{AJ}}                   
\def\araa{\jref@jnl{ARA\&A}}             
\def\apj{\jref@jnl{ApJ}}                 
\def\apjl{\jref@jnl{ApJ}}                
\def\apjs{\jref@jnl{ApJS}}               
\def\ao{\jref@jnl{Appl.~Opt.}}           
\def\apss{\jref@jnl{Ap\&SS}}             
\def\aap{\jref@jnl{A\&A}}                
\def\aapr{\jref@jnl{A\&A~Rev.}}          
\def\aaps{\jref@jnl{A\&AS}}              
\def\azh{\jref@jnl{AZh}}                 
\def\baas{\jref@jnl{BAAS}}               
\def\jrasc{\jref@jnl{JRASC}}             
\def\memras{\jref@jnl{MmRAS}}            
\def\mnras{\jref@jnl{MNRAS}}             
\def\pra{\jref@jnl{Phys.~Rev.~A}}        
\def\prb{\jref@jnl{Phys.~Rev.~B}}        
\def\prc{\jref@jnl{Phys.~Rev.~C}}        
\def\prd{\jref@jnl{Phys.~Rev.~D}}        
\def\pre{\jref@jnl{Phys.~Rev.~E}}        
\def\prl{\jref@jnl{Phys.~Rev.~Lett.}}    
\def\pasp{\jref@jnl{PASP}}               
\def\pasj{\jref@jnl{PASJ}}               
\def\qjras{\jref@jnl{QJRAS}}             
\def\skytel{\jref@jnl{S\&T}}             
\def\solphys{\jref@jnl{Sol.~Phys.}}      
\def\sovast{\jref@jnl{Soviet~Ast.}}      
\def\ssr{\jref@jnl{Space~Sci.~Rev.}}     
\def\zap{\jref@jnl{ZAp}}                 
\def\nat{\jref@jnl{Nature}}              
\def\iaucirc{\jref@jnl{IAU~Circ.}}       
\def\aplett{\jref@jnl{Astrophys.~Lett.}} 
\def\apspr{\jref@jnl{Astrophys.~Space~Phys.~Res.}}
\def\bain{\jref@jnl{Bull.~Astron.~Inst.~Netherlands}} 
\def\fcp{\jref@jnl{Fund.~Cosmic~Phys.}}  
\def\gca{\jref@jnl{Geochim.~Cosmochim.~Acta}}   
\def\grl{\jref@jnl{Geophys.~Res.~Lett.}} 
\def\jcp{\jref@jnl{J.~Chem.~Phys.}}      
\def\jgr{\jref@jnl{J.~Geophys.~Res.}}    
\def\jqsrt{\jref@jnl{J.~Quant.~Spec.~Radiat.~Transf.}}
\def\memsai{\jref@jnl{Mem.~Soc.~Astron.~Italiana}}
\def\nphysa{\jref@jnl{Nucl.~Phys.~A}}   
\def\physrep{\jref@jnl{Phys.~Rep.}}   
\def\physscr{\jref@jnl{Phys.~Scr}}   
\def\planss{\jref@jnl{Planet.~Space~Sci.}}   
\def\procspie{\jref@jnl{Proc.~SPIE}}   

\title[Variations in emission from plasmoid ejecta]{Variations in emission from episodic plasmoid ejecta around black holes}
\author[Z. Younsi and K. Wu]{Ziri Younsi$^{1,2}$\thanks{E-mail: younsi@itp.uni-frankfurt.de (ZY);  \hfill \newline
   kinwah.wu@ucl.ac.uk (KW)} 
and Kinwah Wu$^{2}$\footnotemark[1] \\ \\ 
$^{1}$Institut f\"ur Theoretische Physik, Max-von-Laue-Stra{\ss}e 1, D-60438 Frankfurt am Main, Germany \\
$^{2}$Mullard Space Science Laboratory, University College London, Holmbury St. Mary, Dorking, Surrey, RH5 6NT, UK
}

\begin{document}
\twocolumn

\date{Accepted ***. Received *** in original form ****}

\pagerange{\pageref{firstpage}--\pageref{lastpage}} \pubyear{2015}

\maketitle

\label{firstpage}

\begin{abstract} 

The X-ray and radio flares observed in X-ray binaries and active galactic nuclei (AGN)
 are attributed to energetic electrons in the plasma ejecta 
  from the accretion flows near the black hole in these systems.  
It is argued that magnetic reconnection could occur in the coronae above the accretion disk 
  around the black hole, 
  and that this drives plasmoid outflows resembling the solar coronal mass ejection (CME) phenomenon. 
The X-ray and radio flares are emission from energetic electrons 
  produced in the process. 
As the emission region is located near the black hole event horizon,    
  the flare emission would be subject to special- and general-relativistic effects.    
We present calculations of the flaring emission from plasmoids orbiting around a black hole  
   and plasmoid ejecta launched from the inner accretion disk 
   when general-relativistic effects are crucial in determining the observed time-dependent properties of the emission.  
We consider fully general-relativistic radiative transfer calculations of the emission 
   from evolving ejecta from black hole systems, 
   with proper accounting for differential arrival times 
   of photons emitted from the plasmoids,  
   and determine the emission lightcurves of plasmoids  
   when they are in orbit and when they break free from their magnetic confinement. 
The implications for interpreting time-dependent spectroscopic observations 
  of flaring emission from accreting black holes are discussed. 
\end{abstract}


\section{Introduction} 

Relativistic outflows are characteristic of accreting compact objects,   
  from massive black holes in AGN and quasars 
  to stellar-mass black holes and neutron stars in X-ray binaries. 
The outflows are often intermittent, episodic and transient. 
Knot-like bright features are often seen in the large-scale jets of radio loud AGN, e.g.\ M87 
  \citep[see][]{Reid1989}, 
  which are interpreted as brightened emission from non-thermal electrons 
  freshly accelerated by shocks \citep{Biretta1993,Owen1989,Nakamura2010}  
  that are formed by colliding shells in the jet.  
Superluminal motions of bright features are observed in some quasars, e.g.\ 3C279 \citep{Unwin1989}.  
The association of the superluminal bright features 
  with the variations in the X-ray emission, such as those in the active galaxy 3C120 \citep{Marscher2002},  
  suggest that intermittent outflows could be in the form of plasmoid ejecta.    
Superluminal motions have also been observed in microquasars (black hole X-ray binaries), 
  e.g. GRS~1915+105 \citep{Mirabel1994},   
  GRO~J1655$-$40  \citep[see][]{Tingay1995} and RTXE~J1550$-$564 \citep[see][]{Hannikainen2009}, 
  where individual bright superluminal features were clearly identified in the radio VLBI images.  
In GRS~1915+105 radio flares along with X-ray and IR flares were observed, 
  in particular during state transitions where 
  the X-ray spectrum became softened \citep[see][]{Mirabel1998, Fender1999}. 
The similarity in the temporal properties of the emission from the AGN/quasars and microquasars \citep{Stevens2003} 
  suggests a common driving mechanism for the outflow.    
 The optical depth in the emitting plasmas evolves following a similar trend 
  with only the timescale being stretched or compressed 
  according to the length scale of the system 
  \citep[see e.g. the expanding plasmon model described in][]{vanderLaan1966}.   
Thus, the ejection of plasmoids and the associated flaring in the X-ray and radio wavelengths 
  are universal to accreting black holes across all mass and length scales,  
  and they must occur very close to the black hole event horizon.   

Compared to galactic microquasars and distant quasars, 
  the black hole at the Galactic Centre is relatively dormant.  
 However, sporadic flares in the X-ray, IR and radio wavelengths have been observed in the Sgr A* region, 
  usually as frequently as several episodes per day 
  \citep{Baganoff2001, Genzel2003, Eckart2006, Meyer2008, Neilson2013, Dexter2014, Brinkerink2015}.  
The flare brightness profile generally has a rapid rise and a slow decay.  
Also, the peaks in the lightcurve of different wavelengths show relative delays  \citep{Yusef-Zadeh2006}.  
The flaring properties of the emission from Sgr A* 
  can be explained by the ejection of plasmoids \citep[see][]{Marrone2008},  
  which expand when they emerge from their launching sites. 
The variations in the emission are caused by circulating plasmoids  
  which are anchored above an accretion disk  
  \citep[see e.g.][]{Broderick2005, Meyer2006, Dovciak2006, Noble2007, Dexter2009, Yuan2009, Zamaninasab2011} 
  until they are eventually released.  
  
Ejections of magnetised plasmoids are not unique in accreting black hole systems. 
Multi-component magnetic outflows and ejections have been observed for decades 
  in many other astrophysical systems, including the Sun. 
In the solar environment, the phenomenon is understood 
  in terms of rapid solar winds and coronal mass ejections (CMEs) 
  \citep[see][for reviews]{Forbes2000, Chen2011}.   
Observations aside, 
   there are theoretical reasons 
   as to why the magnetic processes operating in the accretion flow around a black hole 
   and on the solar surface   
   are morphologically and dynamically similar. 
Accretion disks around compact objects are permeated by magnetic fields.
The accretion disk is also enveloped above and below by a low-density, hot and ionised layer, namely the corona, 
  from where Compton X-rays in galactic black hole X-ray binaries and AGN originate. 
These disk coronae are magneto-active and dynamically coupled to strong shearing plasma flows below.   
The interaction is analogous to that between the solar coronae and the solar photosphere.  
Numerical general-relativistic magnetohydrodynamic (GRMHD) simulations of accretion onto black holes 
    \citep[see e.g.][]{DeVilliers2005}
   show flow morphology in accretion disks to be remarkably similar to that near the solar surface.    
Solar CME-like eruption could occur in the magnetised flows around black holes, 
  thus providing a viable model to explain the flaring and episodic plasmoid ejections 
  observed in accreting black hole systems, such as Sgr A* \citep{Yuan2009}.
  
Angular momentum transport in the accretion disks around black holes  
  is regulated by the magneto-rotational instability (MRI) 
  \citep{Balbus2003, BalbusHawley2003}. 
Near the black hole event horizon the dynamics in the accretion flow are complex. 
The flow is highly turbulent, 
  and there is a continuous emergence of magnetic field loops from 
  below the accretion disk surface, injecting magnetic stresses and energy into the corona above. 
With their foot-points anchored onto the disk,  
  the coronal magnetic fields 
  are sheared both by the turbulent motion below and the differential rotation in the accretion disk.  
Magnetic reconnection is inevitable,   
   leading to reconfiguration of the topology and redistribution the helicity of the coronal magnetic fields
  \citep[see][]{Berger1984,Demoulin2007}.  
After reconnection the magnetic fields may continue to evolve and a current sheet is formed. 
When equilibrium breaks down, a plasmoid is expelled.  
Initially, the plasmoid is not completely detached, 
 and it is connected by a current sheet. 
Subsequent magnetic reconnection occurs in the current sheet, 
  creating a strong Lorentz force  
  and accelerating the plasmoid away from the black hole accretion disk.  
GRMHD simulations show that, even under the assumption of a weak initial magnetic field, 
   rapid mass ejections are present, 
   and they are embedded in the continuous background of the jet outflows  
   \citep{Machida2000, DeVilliers2003}.   
The time intervals between successive ejection episodes are typically $\sim1600\, r_{\mathrm{g}}/c$.   
(Here $r_{\mathrm g} = GM/c^2$ is the gravitational radius, $M$ is the black hole mass, 
   $G$ is the gravitational constant, and $c$ is the speed of light.)     

This scenario has been proposed to explain the episodic outflows from the massive nuclear black holes 
   in galaxies, e.g.\ Sgr A* \citep[see][]{Yuan2009}. 
One may ask whether or not the scenario is also applicable 
   for the more dramatic ejection events associated with the radio outbursts of microquasars, 
   e.g.\  GRO~J1655$-$40 \citep[see][]{Tingay1995} and RTXE~J1550$-$564 \citep[see][]{Hannikainen2009}.
To verify the applicability of the CME scenario for plasmoid ejections 
  in accreting black holes across the mass spectrum 
  we must first determine the time-dependent properties of the emission in the plasmoid ejection process. 
As the plasmoids are launched in the regions close to the black hole event horizon, 
  a proper treatment of relativistic and gravitational effects is needed 
  in the radiative transfer calculations. 

In this work we investigate relativistic and general-relativistic effects on the emission 
   from plasmoids with dynamics as described in the CME scenario for episodic jets. 
We carry out radiative transfer calculations 
  and compute the lightcurves from evolving plasmoids 
  with a proper treatment of the arrival time of photons from different regions both on and within the plasmoids, 
  as well as considering the contributions from higher-order lensed photons (which orbit the black hole multiple times before reaching the observer). 
We demonstrate that relativistic effects are important,  
  and that the variations in the emission contain 
  useful information about the system parameters, 
  such as the black hole spin, 
  the geometry of the system 
  and the observer viewing inclination angle with respect to the black hole spin axis. 

The paper is organised as follows.
In \S 2 we summarise the basic equations for relativistic radiation transport calculations in this study 
   and derive the governing equations for both the plasmoid's dynamics and its radiative properties. 
In \S 3 the emission from orbiting plasmoids is calculated and presented in the form of lightcurves.
Different models for the plasmoid emissivity, 
   as well as the effects of photon arrival time in these models,  
   are investigated and the physical consequences are discussed. 
In \S 4 the emission from a magnetically ejected plasmoid is investigated, 
   with consideration given to how different initial configurations 
   will give rise to different morphologies in the emission lightcurves. 
In \S 5 we summarise briefly the key results. 
We also discuss the implications of this work 
  and issues associated with time-dependent properties 
  of emission from objects orbiting around a black hole 
  or ejected from the black hole's vicinity 
  before we present a brief conclusion.

\section{Radiative transfer and plasmoid dynamics} 

\subsection{Space-time metric and radiative transfer equation}

The Schwarzschild radius and gravitational radius of a black hole of mass $M$ 
  are given by $r_{\mathrm s}  =  2M$ and $r_{\mathrm g} = M$ respectively,
  where the natural unit convention $c = G =1$ has been adopted. 
Herein, unless stated otherwise, this convention is maintained throughout.

For a rotating (Kerr) black hole, the space-time interval is given (in Boyer-Lindquist co-ordinates) by 
\begin{eqnarray} 
 {\mathrm d}\tau^2 &=& \left( 1- {{2Mr} \over \Sigma}\right){\mathrm d}t^2 
         + {{4aMr \sin^2 \!\theta} \over \Sigma}\, {\mathrm d}t\,{\mathrm d}\phi 
         - {\Sigma \over \Delta}\, {\mathrm d} r^2 \nonumber \\ 
     & &  - \Sigma {\mathrm d} \theta^2  
    - \left( r^2+a^2 + {{2a^2Mr \sin^2 \!\theta} \over \Sigma} \right) 
                  \sin^2 \!\theta \,  {\mathrm d} \phi^2 ,
\end{eqnarray}   
where  
\begin{eqnarray} 
 \Sigma & \equiv &  r^2+a^2\cos^2 \! \theta \ , \\ 
 \Delta & \equiv & r^2 -2Mr +a^2 \ , 
\end{eqnarray} 
  and $(r,\theta,\phi$) is a three vector in spherical co-ordinates.    
The rotational rate of the black hole is specified by a spin parameter $a$, 
  with  $a/M = 0$ for a Schwarzschild black hole  
  and $a/M = 1$ for a maximally rotating Kerr black hole.        

Electromagnetic radiation propagates along null geodesics of the space-time, 
  and  in the absence of scattering the covariant radiative transfer equation takes the form: 
\begin{eqnarray} 
  \frac{{\mathrm d}{\cal I}}{{\mathrm d}{\lambda}}  
     & = & p^\alpha \frac{\partial {\cal I}}{\partial x^\alpha} 
         + \Gamma^{\alpha}_{\beta \gamma} p^\beta p^\gamma  \frac{\partial {\cal I}}{\partial p^\alpha}  \nonumber   \\  
       & = &  - p^\alpha u_\alpha \big\vert_{\lambda} \left[ - \chi_0 (x^\beta, \nu) {\cal I} + \eta_0 (x^\beta, \nu) \right] \ , \label{GRRT_eqn} 
\label{eq-grrt}   
\end{eqnarray} 
  \citep[see][]{Baschek1997,Fuerst2004,Wu2008},  
   where $\nu$ is the frequency of the radiation,  
   $\cal I$ is the invariant intensity, and    
   $\lambda$ is the affine parameter.  
The absorption coefficient  $\chi_0$ and the emission coefficient $\eta_0$, which are functions of frequency, 
   are evaluated in the local fluid rest frame (hence the subscript $``0"$). 
The transfer equation can be solved by integration along the geodesic 
  using standard ray-tracing methods. 

\subsection{Orbital motion and ejection of the plasmoid} 

For the purpose of this study, 
  we adopt the phenomenological model proposed by \cite{Yuan2009}, 
  which is derived using a solar-CME analogy.  
The development of solar CMEs involve a sequence of processes 
  \citep[see][]{Torok2007,Kliem2010,Chen2011,Janvier2013}. 
If we  ignore the complex details of the MHD, 
  we may divide the plasmoid ejection into two stages: 
(i) the pre-launch stage 
and (ii) the launching stage. 
In the pre-launch stage the plasmoid is anchored to the accretion disk by a magnetic field,  
  and it is orbiting around the black hole above the accretion disk. 
In the launching stage it breaks the magnetic confinement 
  and accelerates upward. 

\subsubsection{Orbital motion of plasmoid anchored to the accretion disk}  

We assume that the plasmoid is magnetised, spherical and filled with energetic electrons. 
It has a radius $r_{\mathrm{p}}=0.5\, r_{\mathrm{g}}$ 
  and is located at a radial distance $r_{\mathrm{c}}$ from the black hole centre. 
The plasmoid sphere is therefore an extended body, leading to several important considerations.  
Firstly, differential rotation may occur within the plasmoid.  
Different parts of the plasmoid will take different times to complete one orbit 
  and the shear will deform the shape of the plasmoid. 
Secondly, there are variations in the gravity across the plasmoid 
  and the emission from the plasmoid is affected by this gravitational field gradient.  
Thirdly, photons emitted from different parts of the plasmoid  
  are lensed differently.  
Different path lengths of the photons lead to different travel times when reaching the observer.   
Thus, photons arriving at the observer's location at the same time 
  in the observer's reference frame   
  may not be emitted at the same time in either the observer's frame 
  or a local reference frame of an emitter in the plasmoid.   

For the plasmoid's shape to remain unchanged throughout an orbit,  
  all parts within the plasmoid must complete an orbit around the black hole 
  in the same time interval as viewed by the observer.  
This requires that the whole plasmoid orbits around the black hole with the same angular velocity 
  as measured by an observer in Boyer-Lindquist co-ordinates. 
We assume that this condition is satisfied 
  and the plasmoid remains spherical throughout its motion. 

As the plasmoid is anchored very close to the accretion disk, 
   as an approximation we may set the plasmoid on the accretion disk in the pre-lauch stage. 
As such the orbital angular velocity of plasmoid's centre around the black hole 
  takes the local Keplerian value, i.e.
  $\Omega_{\mathrm k} = \sqrt{M}\big{/}\left(a\sqrt{M}+{r_{\mathrm{c}}}^{3/2}\right)$.   

\subsubsection{Acceleration of the ejected plasmoid} 

We assume that the plasmoid maintains its shape and density before and after it is launched.  
The model proposed by \cite{Yuan2009} gives a prescription 
  for the vertical (upward) motion of the plasmoid after it is released. 
To first order, the vertical motion is governed by the force equation: 
\begin{eqnarray}
   m\frac{\mathrm{d}^{2}h}{\mathrm{d}t^{2}} & = & \left| \, \mathbf{I}\times\mathbf{B}_{\mathrm{ext}} \right| - F_{\mathrm{g}} \ , \label{plasmoid_height}
\end{eqnarray}
  where $m(t)$ is the total mass within the plasmoid, 
  $h(t)$ is the height of the plasmoid above the accretion disk, 
  $F_{\mathrm{g}}$ is the gravitational force, 
  $\mathbf{I}$ is the integrated current density inside the plasmoid, 
  and $\mathbf{B}_{\mathrm{ext}}$ is the total magnetic field from all sources except $\mathbf{I}$.
The current density $\mathbf{j}$  and magnetic field $\mathbf{B}$ interior to and outside of the plasmoid 
  satisfy the force-free condition (at the zeroth order) 
  $\mathbf{j}\times \mathbf{B} = 0$, with 
\begin{eqnarray} 
  \mathbf{j} &=& \frac{[\,c\,]}{4\pi}\, \nabla \times \mathbf{B} \ .
\end{eqnarray}
To determine the evolution of the plasmoid's velocity, 
  the force equation (equation \ref{plasmoid_height}) must be solved, along with three other parameters. 
In total, five coupled ordinary differential equation are to be solved   
  for the $h$, $\dot{h}$ and $m$ together with two other auxiliary variables ${\bar p}$ and ${\bar q}$.  
  For details, see \citet{Yuan2009}. 

\subsubsection{Evolution of the plasmoid velocity} 

The initial configuration of the plasmoid is the same as that described in \cite{Lin2000}. 
We set the density within the plasmoid to be $\rho_{0}=10^5\:\!{\rm cm}^{-3}$  
  and the initial magnetic field to be $B_0=16\,{\rm G}$, 
  similar to the values used in \cite{Yuan2009}. 
As the density distribution and magnetic field configuration in the accretion disk corona around a black hole 
  are poorly understood, 
  we follow \cite{Yuan2009} and assume that an accretion disk corona is similar to the solar corona. 
The results are not expected to change very much using this assumption or an alternative, 
  provided that the local Alfv\'en speed near the current sheet 
  does not decrease drastically with height 
  within the ejection region of the plasmoid \citep[see][]{Lin2006}.   
This condition is satisfied for a planar geometry as in the corona above an accretion disk. 
 
\begin{figure}
\begin{center} 
\includegraphics[width=0.47\textwidth]{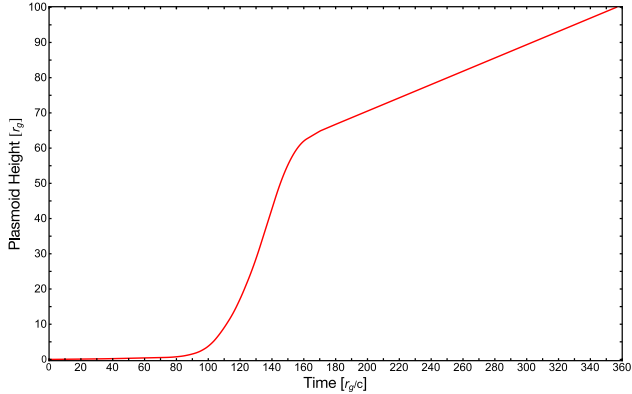} 
\caption{Height of the the magnetically ejected plasmoid above the equatorial plane as a function of time. } 
\label{fig-1}
\end{center}
\end{figure}

Figure \ref{fig-1} shows the evolution of the height of the plasmoid above the equatorial plane,    
  obtained from equation (\ref{plasmoid_height}) 
  with the aforementioned initial conditions for $r_{\mathrm{c}}=5\, r_{\mathrm{g}}$.  
The plasmoid does not show acceleration immediately after its release, 
  and it takes about $100\, r_{\mathrm{g}}/c$ to accelerate from rest to a speed of $0.02\, c$. 
Magnetic reconnection in the form of a huge Lorentz force then feeds the current into the plasmoid, 
  rapidly accelerating the plasmoid to a speed $\gtrsim 0.8\, c$   in $\sim 50\, r_{\mathrm{g}}/c$,
  reaching a height of $\sim 60\, r_{\mathrm{g}}$. 
Thereafter the plasmoid decelerates and sails away with a roughly constant speed.  

In addition to the vertical motion 
  we need to account also for the azimuthal variation of the plasmoid's motion  
  and the changes in its distance from the spin axis of the black hole. 
In realistic situations, turbulence and shocks are present 
  and they could deflect the the vertical path of the plasmoid. 
Also, the plasmoid is expected to follow magnetic field lines 
  that thread into the accretion disk below.  
Modelling turbulence and shocks in the system and their effects, 
  which is non-trivial, is beyond the scope of this work.  
We therefore leave this for future studies and place emphasis on the effects of the magnetic fields.  

\begin{figure}
\begin{center} 
\vspace*{-0.5cm}
\includegraphics[width=0.3\textwidth]{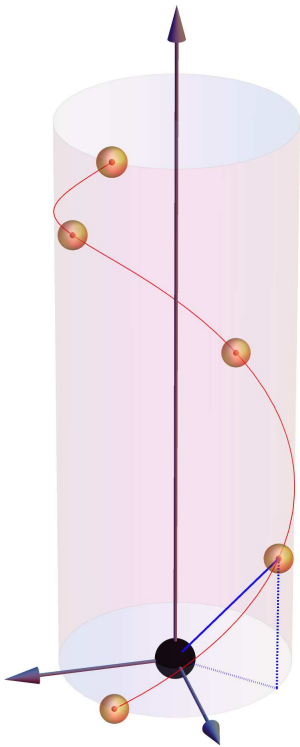} 
\vspace*{-0.3cm}
\caption{Illustration of the time-sequence of the helical motion of a magnetically ejected plasmoid (small spheres)  
   from the vicinity of a black hole (large black sphere).
The plasmoid's trajectory is confined to a cylindrical surface of radius $r_{\mathrm{c}}$ 
   from the spin axis of the black hole (indicated by a dotted line).
} 
\label{fig-2}
\end{center}
\end{figure}

Because of the rotation in the accretion disk, 
  the magnetic fields anchored to it will develop a helical structure, 
hence there will be a helical component to the plasmoid's motion.  
To simplify the modelling of this helical component,  
  we first ignore the plasmoid's linear extent and treat it as a ``big'' particle. 
The plasmoid is initially at rest on the equatorial plane. 
Its cylindrical radial distance to the spin axis that passes through the black hole centre ($r_{\mathrm{c}}$)  
  remains fixed 
  when it travels upward. 
Thus, the motion of the plasmoid is confined to a cylindrical surface  
  (see Figure \ref{fig-2}).

For a particle orbiting around a black hole, 
  the azimuthal component of its velocity 
  is given by $v_{\phi} = {\dot{\phi}}/{\dot{t}}$.
The expressions for $\dot{t}$ and $\dot{\phi}$ for a Kerr black hole can be found in \cite{Fuerst2004}. 
Hence, we have   
\begin{eqnarray}
   v_{\phi}\left(r,\theta\right) 
   &=& \frac{\mathrm{cosec} \ \!\theta\sqrt{M(2r^{2}-\Sigma)}}   {\Sigma\sqrt{r}+a\sin\theta\sqrt{M(2r^{2}-\Sigma)}} \ .
\end{eqnarray}
The confinement to a fixed cylindrical surface with $r_{\mathrm{c}}$ 
  implies that the plasmoid's velocity may take the following parameterised form as a function of $r$:
\begin{eqnarray}
  v_{\phi}(r) & =& 
  \frac{r^{2}}{r_{\mathrm{c}}}
  \left\{ \sqrt{\frac{r}{M}}\left[ \frac{r^{4}+a^{2}\left(r^{2}-r_{\mathrm{c}}^{2}\right)}{\sqrt{r^{4}-a^{2}\left(r^{2}-r_{\mathrm{c}}^{2}\right)}} \right] + a\, r_{\mathrm{c}} \right\}^{-1} . \label{v_phi_plas}
\end{eqnarray}
This expression can be in turn parameterised as a function of time 
  by introducing the plasmoid height, $h(t)$: 
\begin{eqnarray}
v_{\phi}(t) &=& \frac{R(t)^{2}}{r_{c}}\left[ \sqrt{\frac{R(t)}{M}}\frac{R(t)^{2}+a^{2}h(t)^{2}}{\sqrt{R(t)^{2}-a^{2}h(t)^{2}}} 
 + a\, r_{\mathrm{c}} \right]^{-1} , \label{v_phi_plas_t}
\end{eqnarray}
   where
\begin{eqnarray}
R(t)^{2} & \equiv & r_{\mathrm{c}}^{2} + h(t)^{2} \ .
\end{eqnarray}

Hereafter we set $M=1$, which is equivalent to normalising the length 
  to $r_{\mathrm{g}}$ and the time to $r_{\mathrm{g}}/c$. 
The change in azimuth for the plasmoid may now be calculated 
   by integrating equation (\ref{v_phi_plas_t}) between its initial and final position: 
\begin{eqnarray}
   \Delta \phi & =&  \int_{t_{i}}^{t_{i+1}}\mathrm{d}\:\!t \ \! v_{\phi} \!\left( t \right) \ .
   \label{delta_phi_plasmon}
\end{eqnarray} 

\begin{figure}
\begin{center} 
\includegraphics[width=0.47\textwidth]{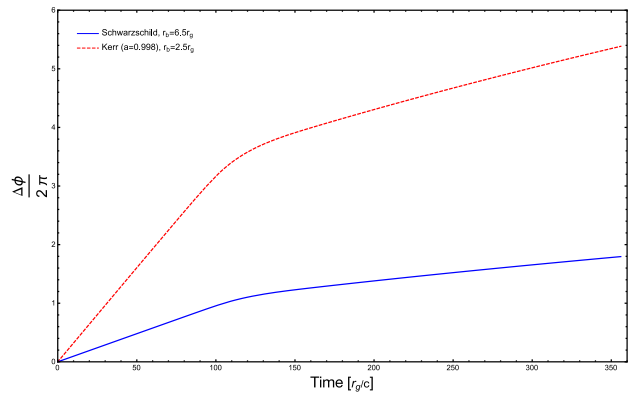} 
\caption{Change in azimuth of the plasmoid 
  obtained from equation (\ref{delta_phi_plasmon}) 
  for the case of a plasmoid ejected at $r_{\mathrm{c}}=6.5\,r_{\mathrm{g}}$ 
  from a Schwarzschild black hole (solid curve) 
  and for the case of a plasmoid ejected at $r_{\mathrm{c}}=2.5\,r_{\mathrm{g}}$ from a  Kerr black hole 
  with $a=0.998$ (dashed curve). 
} 
\label{fig-3}
\end{center}
\end{figure}

A particle orbiting in the equatorial plane has a velocity $v_{\phi} \propto (a+r^{3/2})^{-1}$.  
Thus, the initial location of the plasmoid  (i.e.\ $r_{\mathrm{c}}$) 
  rather than the spin of the black hole 
  will play a more important role in determining the azimuthal evolution $\Delta \phi$. 
Figure \ref{fig-3} illustrates the change in the azimuthal location of the plasmoid 
  for cases with $a=0$, $r_{\mathrm{c}}=6.5\,r_{\mathrm{g}}$ 
  and $a=0.998$, $r_{\mathrm{c}}=2.5\,r_{\mathrm{g}}$. 
As shown, the change is larger for the case of the Kerr black hole, 
  which allows a smaller $r_{\mathrm{c}}$. 
The relative change in the two cases can also be understood as follows. 
Plasmoids ejected closer to the black hole event horizon
  experience much stronger gravitational forces 
  and perform more orbits in a given time than their more distantly ejected counterparts.    

\subsection{Emissivity of the plasmoid}

The emission from the plasmoid depends on the radiative processes and the optical depth across it. 
A fully self-consistent calculation of the emission requires 
  solving the transport equations for the acceleration and the radiative and non-radiative energy loss  
  of the charged particles in the magnetised plasmoid under the influence of the black hole's extreme gravity.  
This is a problem which warrants a proper separate study. 
In this study, we use a phenomenological approach to illustrate 
  the time-dependent properties of the emission from the plasmoid.  
We characterise the plasmoid by its effective optical depth. 
The two representative cases are  
  (i) an opaque plasmoid,   
  and (ii) an optically thin plasmoid that emits synchrotron radiation.  

For the opaque plasmoid, only its surface contributes to the radiation seen by a distant observer. 
In our calculation we assume 
  a local specific surface emissivity following a Gaussian profile for the opaque plasmoid: 
\begin{eqnarray}
    j_{\nu}\left(z,t \right) &  \propto &  \exp \left\{ -\frac{|z-z_{\mathrm{p}}(t) |^{2}}{r_{\mathrm{p}}^2} \right\} \ ,
\end{eqnarray}
   where  $\nu$ is the frequency of the emission,  
   and $z_{\mathrm{p}}(t)$ is the plasmoid position at time $t$ 
  (with $z\equiv r \cos\theta$ in Boyer-Lindquist co-ordinates).  
The results actually do not change much 
   by assuming this surface emissivity profile or an alternative (such as a uniform surface emissivity) 
   for the plasmoid,  
   as dynamical and relativistic effects are more important 
   in determining the time-dependent properties of the radiation emitted from it.    

For an optically thin plasmoid, 
  the emission from all parts within it will reach the observer, 
  and the emission from different location is subject to different relativistic effects.  
In our calculation, we first generate an ensemble of ``emitters'' inside the plasmoid sphere 
  and then sum the emission from these emitters,   
  with the corrections for the frequency shifts and Lorentz intensity boosts,   
  using radiative transfer calculations as described in \citet{Fuerst2007} and \citet{Younsi2012}.  

We consider a Gaussian profile for the density distribution of the non-thermal electrons in the plasmoid, i.e.\   
\begin{eqnarray}
      \rho(\mathbf{x},t) & = & 
      \rho_{\mathrm{c}} \exp \left\{ -\frac{|\mathbf{x}-\mathbf{x}_{\mathrm{p}}\left(t\right)\! |^{2}}{\mathcal{C}^{2}} \right\} \ ,
\end{eqnarray}
   with 
\begin{eqnarray}
\mathcal{C} & = & \frac{r_{\mathrm{p}}}{\sqrt{\ln \left( \rho_{\mathrm{c}}/ \rho_{\mathrm{min}}\right)} } \ ,
\end{eqnarray}
    where $\rho_{\mathrm{c}}$ is the central plasmoid electron number density 
    and $\rho_{\mathrm{min}}$ is the electron number density on the surface of the plasmoid. 

The non-thermal electrons emit radio synchrotron radiation and the plasmoid is optically thin to this radiation.  
The electron synchrotron emissivity may be expressed as   
\begin{eqnarray}  
   j_\nu(\mathbf{x},t)  & = & 
      \rho(\mathbf{x},t) \ \! \mathcal{K}\!\left(p\right) N_{0} \ \! B^{\:\!(p+1)/2} \left(\frac{\nu}{c_{1}} \right)^{(1-p)/2} \ , 
\end{eqnarray} 
  \citep[see][]{Ginzburg1965,Pacholczyk1970},  where
\begin{eqnarray}
\mathcal{K}\!\left( p \right) & = & \frac{c_{2}}{4} 
   \left( \frac{p+\frac{7}{3}}{p+1}\right)\frac{\Gamma\left(\frac{3p-1}{12} \right)\Gamma \left(\frac{3p+7}{12} \right)
     \Gamma \left(\frac{p+5}{4} \right)}{\Gamma\left(\frac{p+7}{4} \right)} \ ,
\end{eqnarray}
   with constants $c_{1}$ and $c_{2}$  given by
\begin{eqnarray}
  c_{1} & = & \frac{3\ \![\,c\,]}{2\pi}\frac{e}{E_{\mathrm{e}}^{3}} \ , \\
  c_{2} & = & \frac{1}{8}\sqrt{\frac{3}{\pi}}\frac{e^{3}}{E_{\mathrm{e}}} \ ,
\end{eqnarray}
Here $\Gamma (...)$ denotes the Gamma function, 
  $E_{\mathrm{e}}=m_{\rm e}[\,c\,]^{2}$ is the electron rest mass energy, 
  $m_{\rm e}$ is the electron rest mass and 
  $e$ is the electron charge. 
The energy distribution of the electrons, $N(E)$, is a power law with spectral index $p$, i.e.\   
  ${N}(E)={N}_{0}E^{-p}$, 
  where ${N}_{0}$ is the normalisation constant determined by the total number density of the electrons.   
 
As we have not modelled the magnetic reconnection process and its associated particle acceleration explicitly,  
 the amount of non-thermal electrons in the plasmoid is an uncertain parameter in the calculations.  
To determine the normalisation $N_0$, 
  we assume energy equipartition between the energetic electrons and the magnetic field.  
From this we obtain    
\begin{equation}
  {N}_{0} = \frac{B^{2}}{8\pi} \times \begin{cases}
    \left[\ln \left(E_{\mathrm{max}}/E_{\mathrm{min}} \right) \right]^{-1} \ , & \text{if $p=2$} \\
    (2-p)\left(E_{\mathrm{max}}^{2-p}-E_{\mathrm{min}}^{2-p}\right)^{-1} \ , & \text{otherwise} \ .
  \end{cases}
\end{equation}
Here $E_{\mathrm{max}}/E_{\mathrm{e}}=\gamma_{\mathrm{max}}$, 
  with $\gamma_{\mathrm{max}}$ denoting the maximum Lorentz factor of the plasmoid electrons 
  (and equivalently for $\gamma_{\mathrm{min}}$). 
  
In all calculations the magnetic field in the plasmoid is taken to be 
  $B=16\,{\rm G}$, similar to the value used in \cite{Yuan2009}. 
The other values for the plasmoid parameters are: 
   $\rho_{\mathrm{c}}=10^{5}\mathrm{cm}^{-3}$, 
   $\rho_{\mathrm{min}}=10^{-5}\mathrm{cm}^{-3}$,  
   $\gamma_{\mathrm{max}}=10^{10}$,  
   $\gamma_{\mathrm{min}}=1$ and 
   $p=2.3$. 

\subsection{Time-keeping in the lightcurve construction}  

The plasmoid is anchored onto an accretion disk before it is released, 
  and the variations in the emission from the system are mainly due to the plasmoid's motion, 
  in particular the orbital circulation. 
On the temporal evolutionary timescale of the plasmoid's motion (before and after its release), 
 the accretion disk is relatively stationary 
 and its emission is practically a uniform background continuum.   
We may therefore ignore the disk emission without loss of generality  
  and consider only the plasmoid's emission. 

We use a ray-tracing method for radiative transfer, as in \cite{Younsi2012}, 
  to generate emission images of the plasmoid.  
The lightcurves are derived from time sequences of the plasmoid images. 
Conventionally,  
  one may sum the emission flux, within a particular frequency band, 
  over the pixels associated with the source in the images following a time sequence 
  to construct a lightcurve of an emission source in that frequency band. 
This approach is inapplicable for time-dependent relativistic systems  
  such as the orbiting plasmoid around a black hole considered here,  
  where the dynamical timescale of the plasmoid    
  is comparable to the light-crossing time across the plasmoid's orbit  
  and the emission is strongly lensed by the black hole's gravity.  
The differences in the propagation times of photons  
  (measured in the observer's frame) 
  emitted from different parts of the system   
  will convolve with relativistic effects, 
  such as Doppler shift and time-dilation of the radiation (photons) and 
  Lorentz boosts of the radiation intensity,  
  leading to variations in the specific flux density. 
These effects not only add complexities to the lightcurve 
  but also mask the variations intrinsic to time-dependent radiative or dynamical processes of the system.  
Also, echoes that are produced by the delayed arrival of the high-order lensed radiation   
  contaminate the signals of the direct emission from the plasmoid at the later part of the orbit.   
Photons arriving at a particular time registered by the observer 
  may spread across a number of images in the time sequence 
  and each image contains photons that have circulated around the black hole many times.
This coupled with photons which propagate away from the plasmoid without completing an orbit around the black hole  
 means proper time-keeping is essential in both ray-tracing and image generation 
  so as to enable the correct extraction of photons for the lightcurve.  
  
Within the calculations themselves there are subtle issues regarding the time-keeping, 
  as the orbital motion of the plasmoid is expressed in terms of coordinate time  
  while in the lightcurve 
  the emission flux is measured by the observer with a clock defined in its local reference frame.   
We consider the following procedures to relate the two reference frames 
   and correct for the time differences. 
We first calculate the plasmoid's trajectory  
  and tabulate the plasmoid's location in the coordinate time. 
Next we define a full orbital phase as one orbital period of the plasmoid as seen by the observer 
  and divide this orbital phase into 100 equal intervals, 
  with equal time widths as measured in the observer's reference frame. 
The lightcurve is a time sequence of emission flux measured in the observer's reference frame 
  as a function of the orbital phase defined above.   
It is constructed from the plasmoid images    
  with the associated time delay in the arrival of each photon 
  subtracted from the observer's local time.  

We compute 100 images for all of the cases that we consider. 
The images are equally spaced in the co-ordinate time throughout the temporal evolution of the plasmoid. 
Each image is calculated from a ray-tracing of $10^6$ photons and   
each photon within an image is identified by its frequency (energy shift), arrival time 
  and pixel location on the image plane. 
For each pixel, after we have sorted the photons into 1000 bins according to their frequencies, 
  we consider the following time ordering procedure for the photons 
  in the images across the image time sequence.    
(i) We first sort the photons in each pixel and each frequency bin 
    by their arrival times, from first to last.  
(ii) We create 100 equally-spaced time bins 
  and determine the total specific emission flux (from the photon counts) 
  for each of these time bins.   
(iii) We then order the specific emission fluxes in the 100 time bins to make a time sequence, 
  and obtain the arrival time-corrected time-dependent spectrum at each pixel. 
The total emission flux at each pixel is obtained by 
   an integration of the specific emission flux 
   over the frequency band (all evaluated in the observer's reference frame) at the pixel.   
Summing the fluxes of all pixels in the image time sequence 
  following the time progression gives the integrated flux lightcurve.  
Steps (i) and (ii) are omitted in the construction of the lightcurves 
  without proper correction of the arrival time of the photons 
  shown in this study.  
Finally, a centred 5-point Gaussian smoothing is used 
  to remove the small noise fluctuations in the lightcurve  
  that are caused by the finite number of photons in the calculations. 
   

\section{Emission from an orbiting plasmoid} 

We consider two cases as a demonstration: 
   (i) an opaque plamoid with only its surface contributing to the radiation, 
  and (ii) an optically thin plasmoid where synchrotron radiation emitted 
  from all parts within it is visible.   
Effects due to differential arrival time are significant only  
  when the plasmoid and the light paths of its emission are sufficiently close 
  to the black hole event horizon, 
  and in this situation arrival time correction is needed.    
For brevity, the arrival time-corrected lightcurve calculations 
  are presented only for the orbiting opaque plasmoid.   
The results obtained in this case hold for other cases,  
 and will be discussed in more detail in \S~\ref{sec-TCLC}.

Calculations are shown for black holes with two different spins.  
The first case places a plasmoid in orbit around a Schwarzschild black hole (i.e.\ $a =0$), 
  at a distance of $r_{\mathrm{c}} = 6.5\, r_{\mathrm{g}}$\footnote{This distance is chosen so that the inner radius of the plasmoid coincides with the innermost stable circular orbit (ISCO)}, i.e. at the ISCO. 
The second case places a plasmoid in orbit around a Kerr black hole with spin parameter $a=0.998$   
   at a distance of $r_{\mathrm{c}} = 2.5\, r_\mathrm{g}$, 
   i.e.\ within $0.8\, r_{\mathrm{g}}$ of the ISCO. 
The plasmoid orbits in an anticlockwise direction when viewed from above,    
   and it is in prograde orbit with respect to the Kerr black hole's rotation.  
The azimuthal position in the plasmoid's orbit is defined such that  
  $\phi = 0^\circ$ corresponds to the plasmoid being in front of the black hole 
  with respect to the observer in the co-ordinate frame, 
  and $\phi = 180^\circ$ to the plasmoid being behind the black hole.    
We use the following convention for the phases in the plasmoid's lightcurve.  
The phase progresses linearly with time measured in the observer's reference frame. 
Phase 0 in the lightcurve corresponds to the location of the plasmoid's centre at $\phi = 90^\circ$,  
  where the emission from that location is expected 
  to have the strongest relativistic Doppler redshift (in the Schwarschild case)
  due to the plasmoid's orbital motion.   

\begin{figure}
\begin{center} 
\vspace*{0.2cm}
  \includegraphics[width=0.23\textwidth]{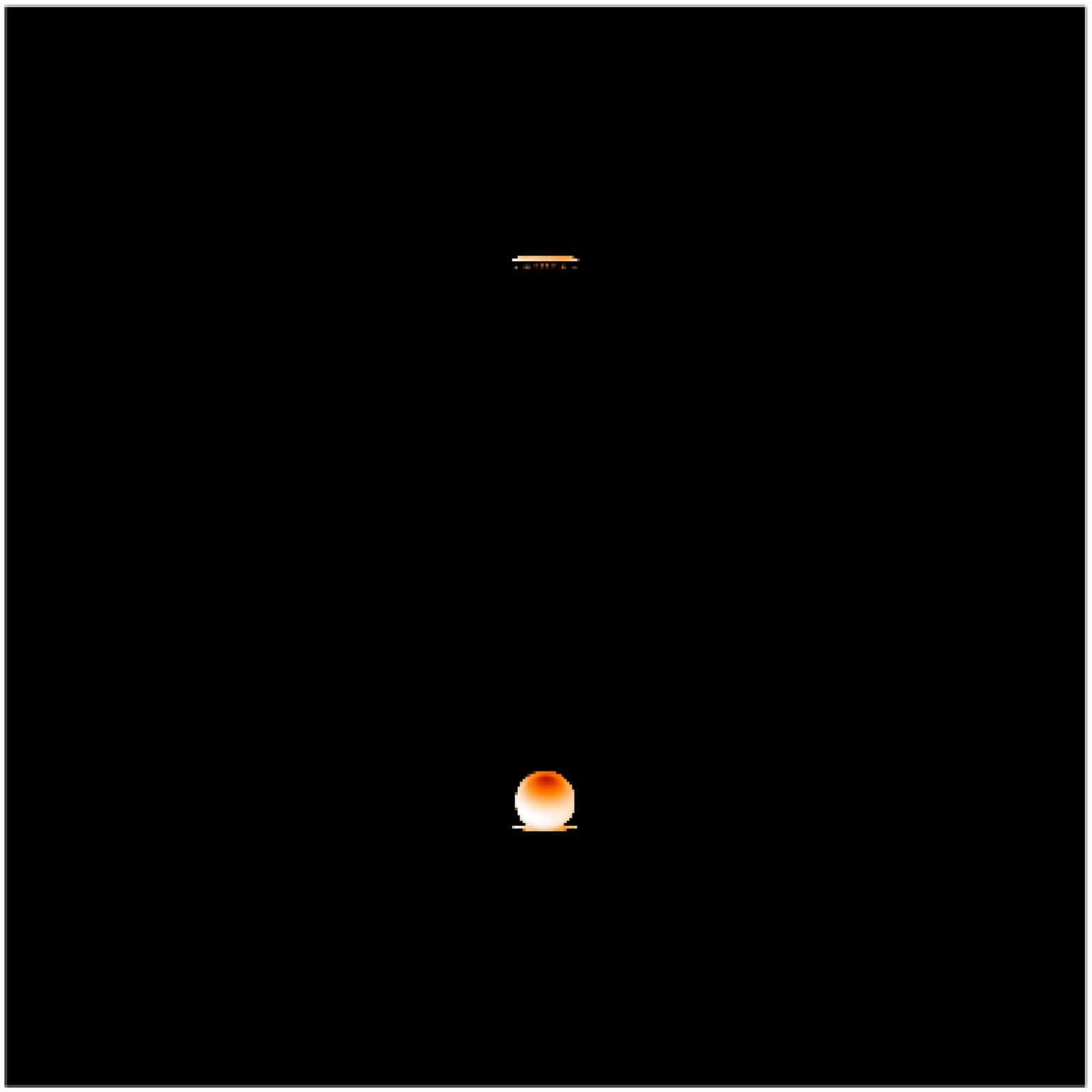}\hspace{-0.825mm}
  \includegraphics[width=0.23\textwidth]{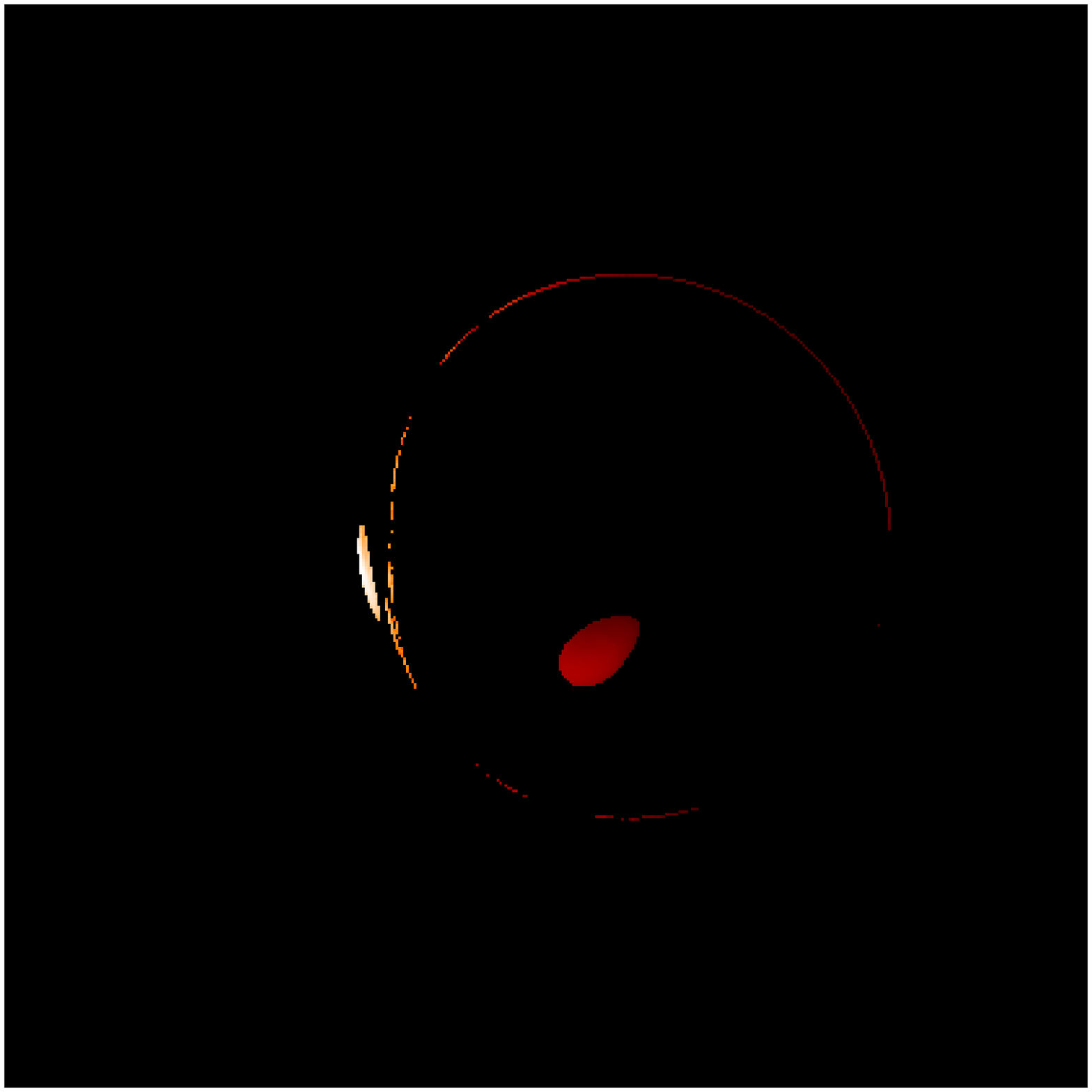}\vspace{0mm}
  \\
  \includegraphics[width=0.23\textwidth]{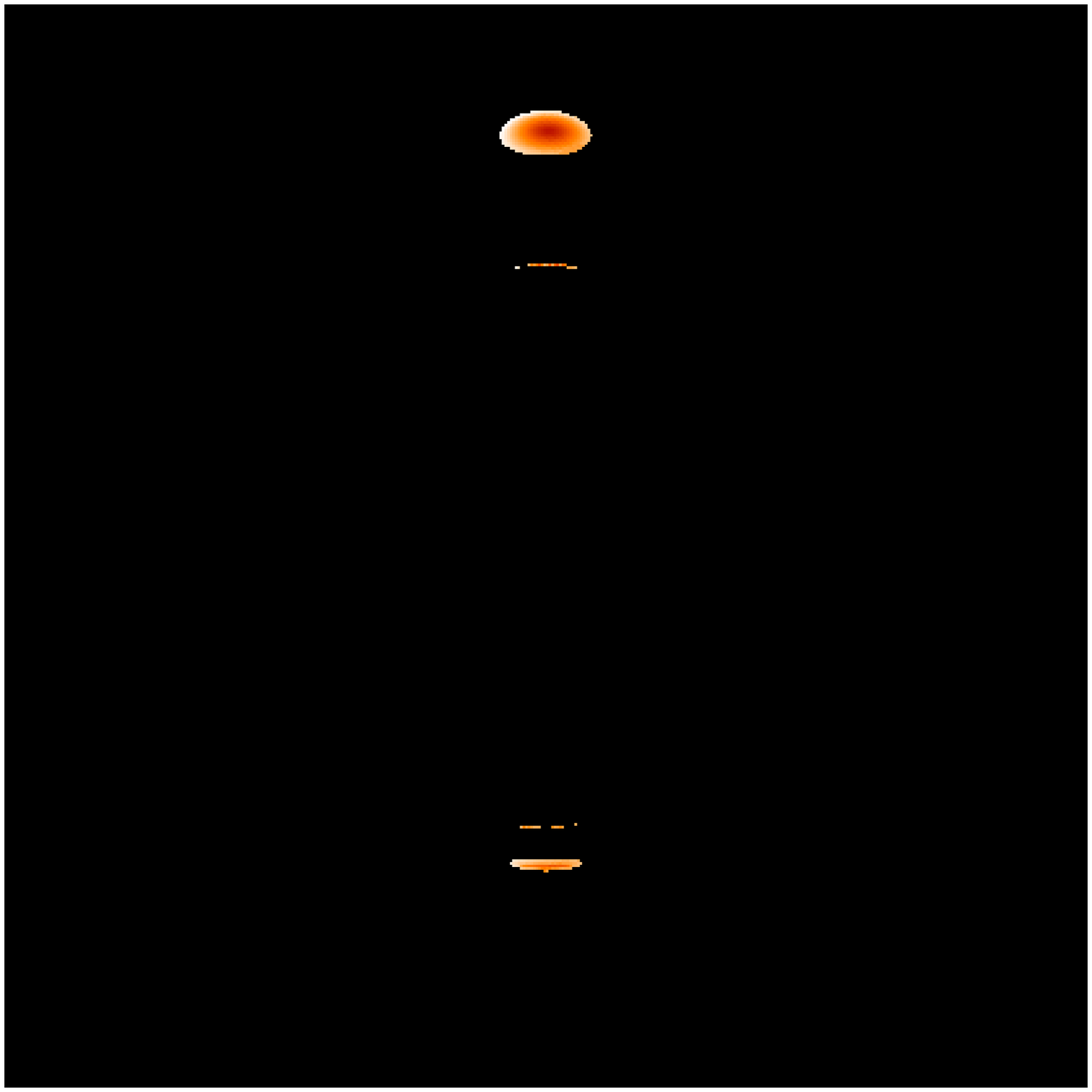}\hspace{-0.825mm}
  \includegraphics[width=0.23\textwidth]{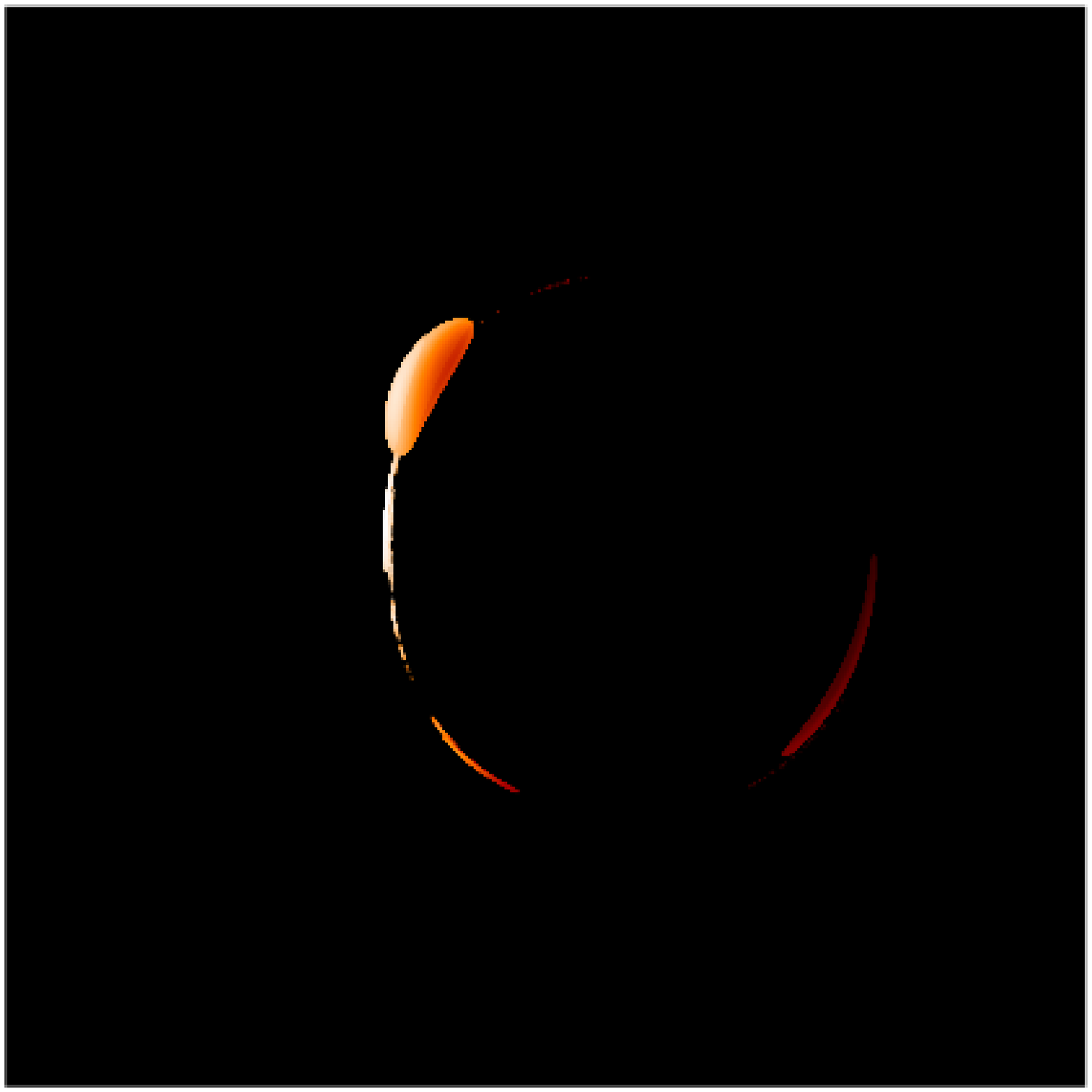}\vspace{0mm}
\caption{Snapshot images of an opaque plasmoid orbiting a Schwarzschild black hole (left) 
   and a Kerr black hole with $a=0.998$ (right),  
   viewed at an inclination angle $i=45^{\circ}$. 
The orbital rotation is in the $e_{\phi}$-direction, 
  i.e.\ anticlockwise as viewed from above. 
From top to bottom: the plasmoid is located at $\phi=0^{\circ}$ 
    (in front of the black hole) and $\phi=180^{\circ}$ (behind the black hole), respectively. 
Arcs of higher-order images of emission from the the plasmoid 
   that have orbited around the black hole multiple times 
   before reaching the observer are clearly visible.
  } 
\label{fig-4}
\end{center}
\end{figure} 

\begin{figure}
\begin{center} 
\vspace*{0.2cm}
  \includegraphics[width=0.23\textwidth]{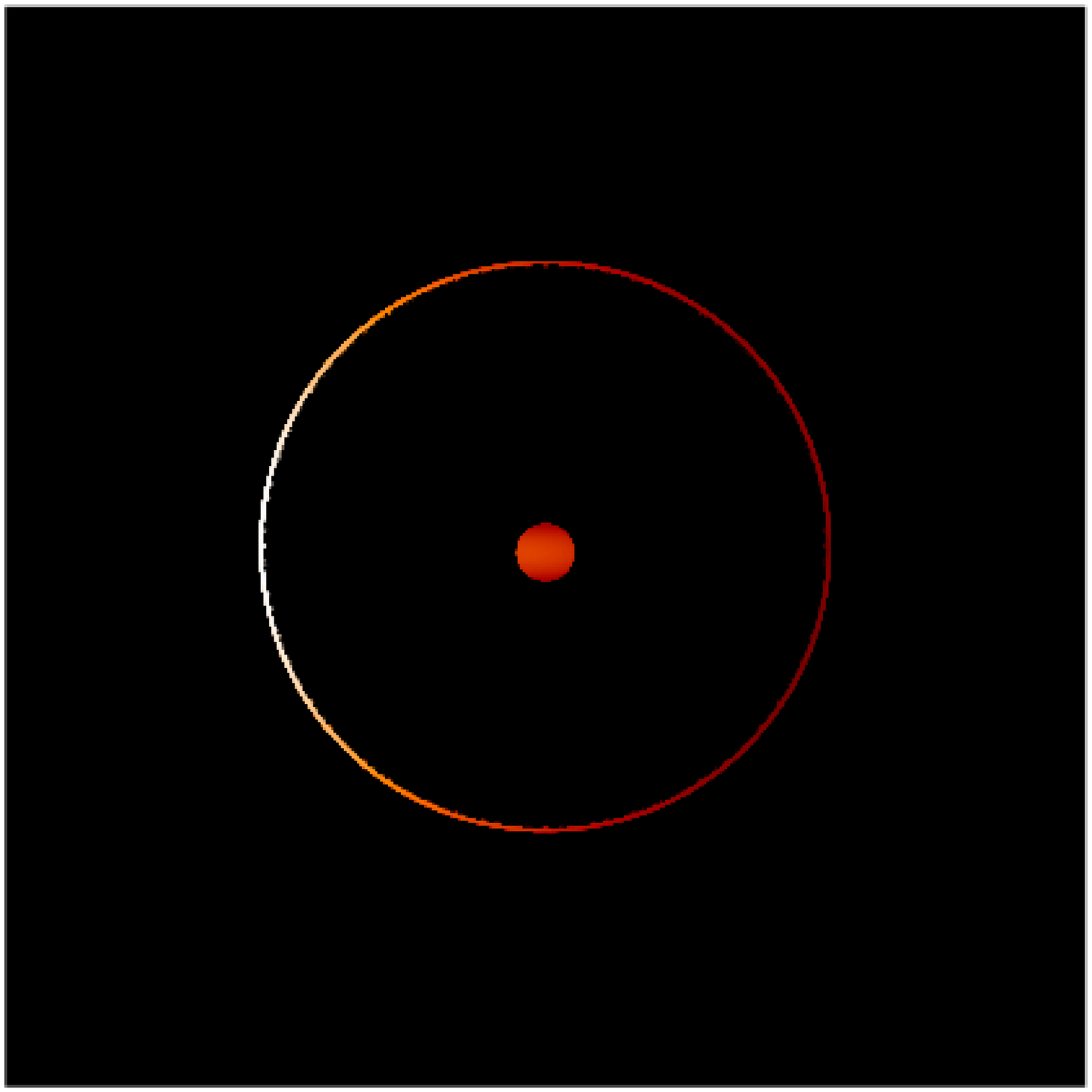}\hspace{-0.825mm}
  \includegraphics[width=0.23\textwidth]{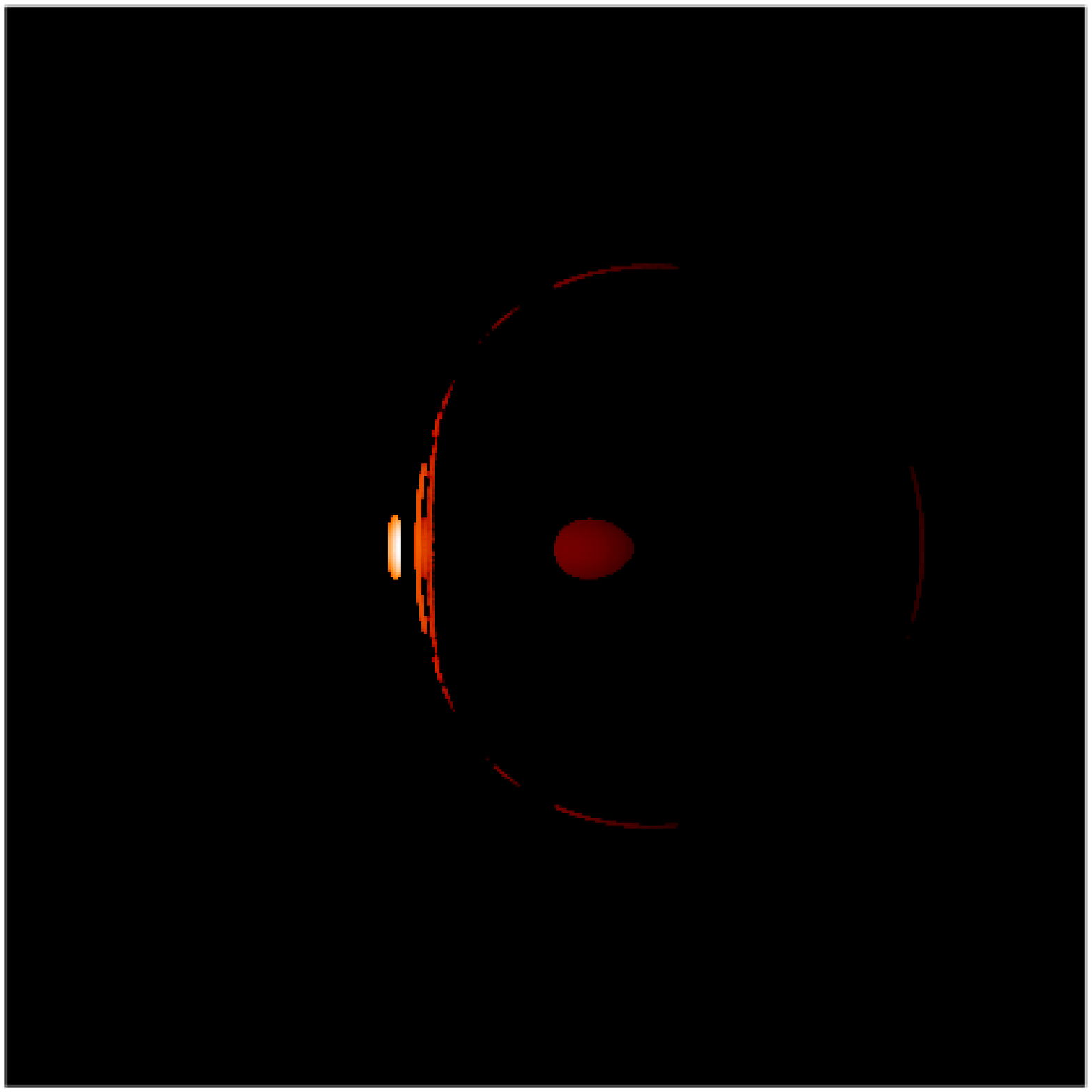}\vspace{0mm}
  \\
  \includegraphics[width=0.23\textwidth]{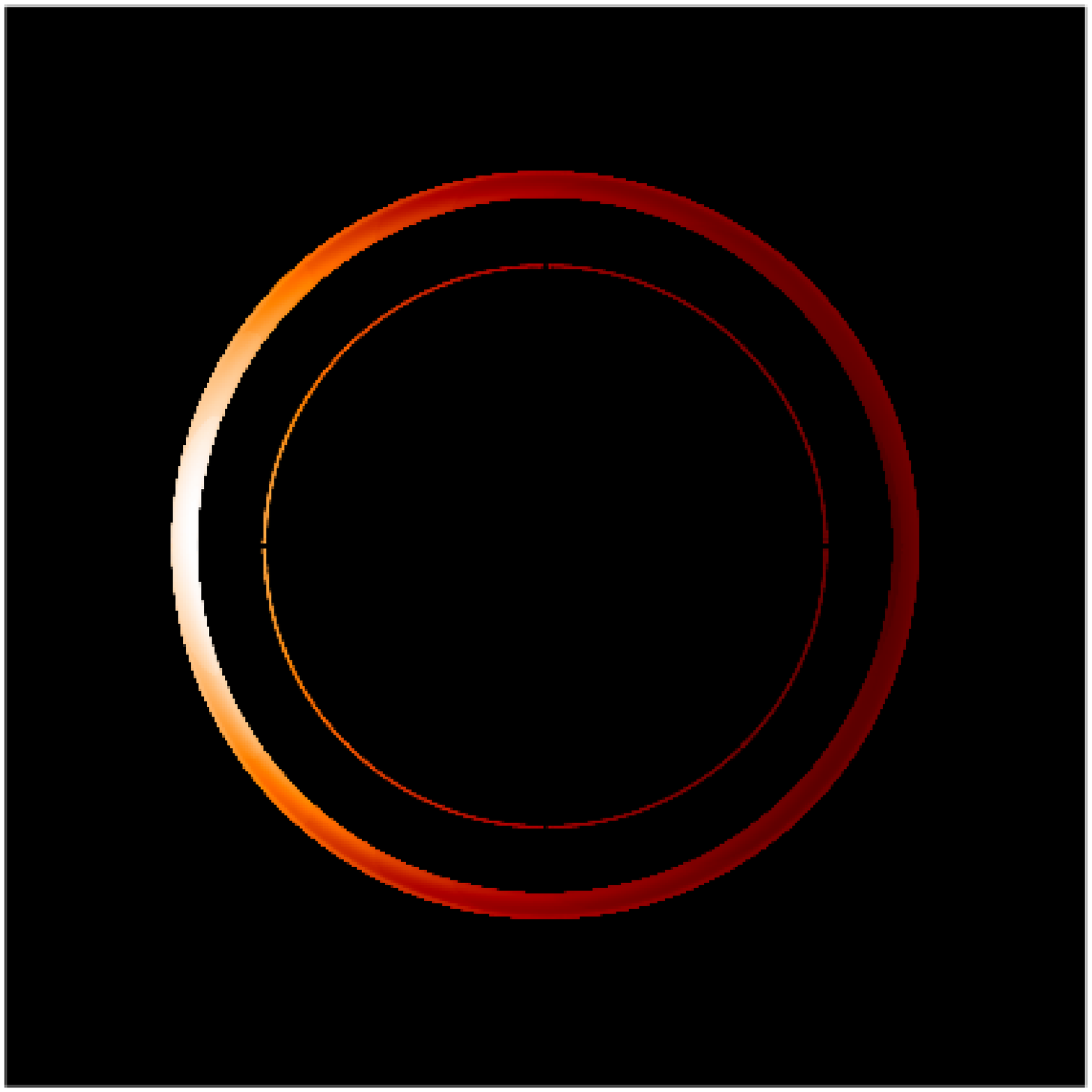}\hspace{-0.825mm}
  \includegraphics[width=0.23\textwidth]{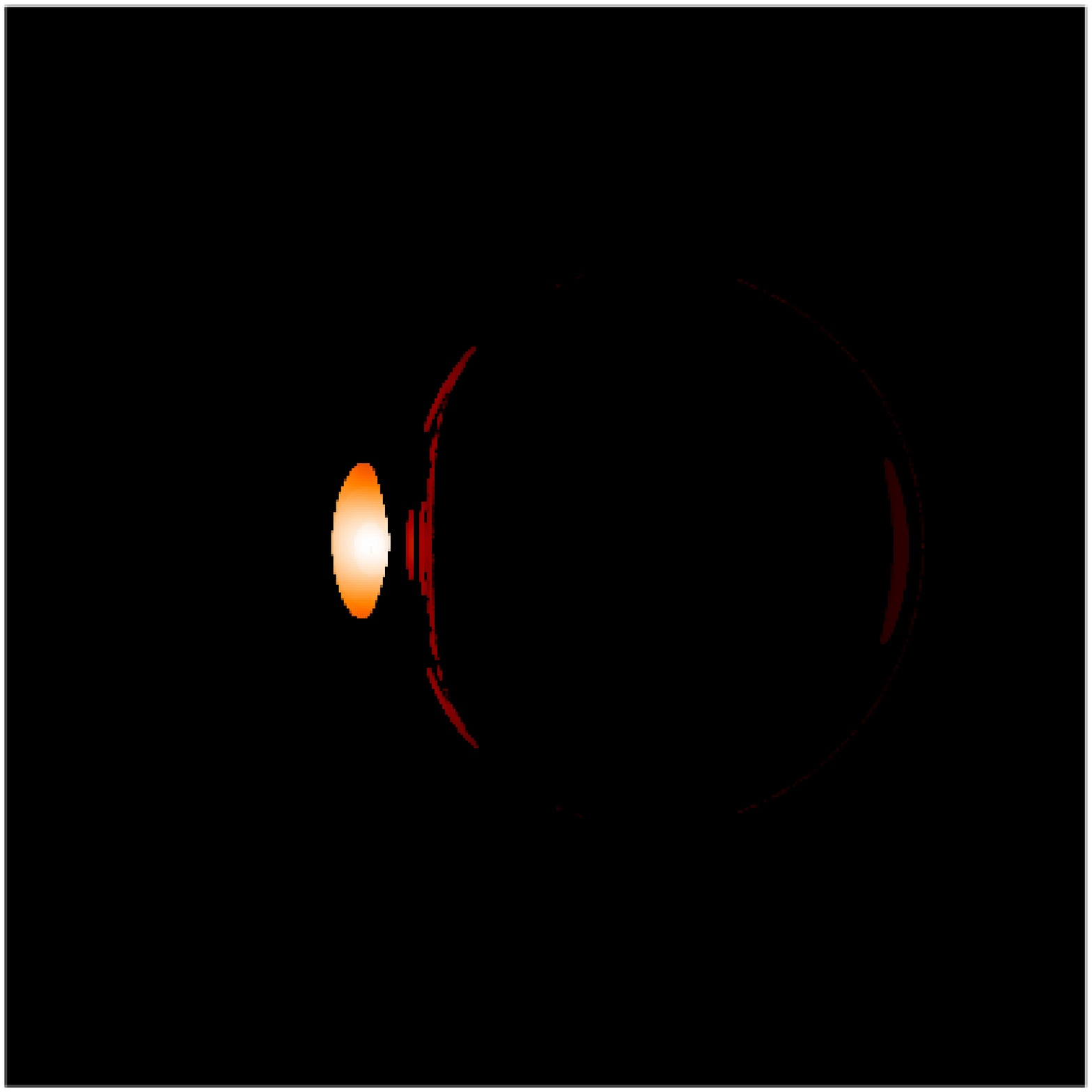}\vspace{0mm}
\caption{As in Figure \ref{fig-4}, but for a viewing inclination angle of $i=89^{\circ}$. 
Note the prominent circular Einstein ring formed 
  when the plasmoid is directly behind the Schwarzschild black hole (bottom left panel).  
}
\label{fig-5}
\end{center}
\end{figure}

\subsection{Opaque plasmoid}

The ray-traced images of an opaque plasmoid orbiting a Schwarzschild black hole 
   and a Kerr black hole with $a=0.998$  
   are shown in Figures \ref{fig-4} and \ref{fig-5} 
   (for viewing inclination angles $i = 45^\circ$ and $89^\circ$ respectively).  
Time correction has not been applied in generating the images in the Figures. 
At $i = 45^\circ$, the plasmoid appears to spherical without noticeable visual deformation in the direct image 
  when it is in front of the Schwarzschild black hole (at $\phi = 0^\circ$). 
The first-order (primary) image consists of the direct image 
  and an image formed by the lensed emission from the back of the plasmoid  
  (which is otherwise invisible) lensed into view.  
The latter part of the first-order image, which is located above the direct image in the image plane, 
  resembles a thin bar.  
The next higher-order image, which is dim and barely visible in the figure, 
  is a small thin bar just below the direct image.     
The images become progressively dimmer with increasing image order.   
When the plasmoid is located on the far side behind the black hole (at $\phi = 180^\circ$),   
  it is flattened in the direct image,  
  and other images of all orders resemble small arcs. 
The images become dimmer as the image order increases.   

At the plasmoid's location, $6.5\, r_{\mathrm{g}}$ from the centre of the Schwarzschild black hole,  
  gravitational effects are substantial but not too severe.   
When viewed at $i = 45^\circ$, a moderate inclination,  
  the emission from the plasmoid is not severely lensed, 
  and the plasmoid preserves a recognisable spheroidal shape, 
  at least in the direct image, 
  throughout most of its orbit.   
In comparison, the appearance of the plasmoid that orbits a Kerr black hole (see Figure \ref{fig-5})  
   is more visually distorted from its intrinsic spherical shape 
   throughout all the image orders 
   even at moderate viewing inclinations. 
This is in part due to the location of the plasmoid closer to the black hole, 
  at $r = 2.5\, r_{\mathrm{g}}$,  
  where stronger gravity gives rise to more prominent lensing effects, 
  and also due to the rotational frame dragging caused by the spinning black hole.  
Rotational frame dragging, together with gravitational lensing,  
  not only causes the plasmoid's direct image to deviate from its spherical shape   
  but also induces asymmetries in the plasmoid's image at all image orders.  
The plasmoid images are elongated along the direction of the black hole's rotation. 
At $ i = 45^\circ$, the plasmoid still maintains a shape resembling a filled-volume object in the direct image. 
It is spatially spread in the lensed images at all image orders,  
   where some form long arcs which almost close up into a loop.  
Comparing the images of the plasmoid at $\phi = 0^\circ$ (directly in front of the black hole) 
   and $\phi = 180^\circ$ (directly behind the black hole) 
   reveals that the convolution of gravitational lensing, rotational frame dragging 
   and viewing geometry of the system is complicated.  
When the plasmoid is behind the black hole,   
  it is not just simply lensed into view like the plasmoid orbiting a Schwarzschild black hole.  
Its emission is subject to rotational frame dragging by the black hole, 
  leading to complex morphological progression when the image order increases. 
The apparent brightness of the plasmoid in different image orders 
  is determined by the competition between brightening due to beaming 
  (caused by gravitational lensing and rotational frame dragging) and 
  dimming due to time-dilation (caused by gravitational redshift and transverse Doppler shift), 
  whereas complicated convolution of gravitational lensing with rotational frame dragging 
  is absent if the black hole is not rotating. 
As a result, the dimming of the plasmoid images does not always progress in a monotonic manner 
  with the increase of the image order (see Figure \ref{fig-4}, top right panel), 
  in contrast to that seen in the case of the Schwarzschild black hole. 

Lensing effects are expected to be prominent for high viewing inclination angles.  
At $i = 90^\circ$,  while the direct image of a plasmoid 
  in front of a Schwarzschild black hole  (at $\phi = 0^\circ$) is spherical,  
  all lensed images, regardless of the image order, are concentric circular rings.   
The lensed images become dimmer as the image order increases. 
When the plasmoid is behind the black hole (at $\phi = 180^\circ$)    
   it is lensed into concentric circular rings (the Einstein rings) for all image orders. 
The outermost thick ring is the first-order image, which is the brightest. 

When the plasmoid is placed closer to the black hole, in the case of the Kerr black hole
  the radiation is subject to stronger suppression by gravitational redshift (time-dilation effect). 
The plasmoid images are therefore dimmer than those in the case of the Schwarzschild black hole.  
 At viewing inclinations close to $90^\circ$, 
  the direct image of a plasmoid in front of a Kerr black hole (at $\phi = 0^\circ$) is slightly elongated, 
  a consequence of rotational frame dragging by the black hole.    
This direct image is enclosed inside asymmetric closed loops of the lensed images. 
Note that the direct image is not always brighter than the lensed images (see Figure \ref{fig-5}),  
  because rotational frame dragging can bring the relativistically boosted emission 
  beamed in the forward direction along the plasmoid's motion into the observer's line-of-sight.  

Rotational frame dragging and gravitational lensing together 
  can cause various optical illusions in the ray-tracing images.   
For instance the first-order lensed image 
  can appear to be dragged outside the loops of higher-order images 
  by the black hole's rotation 
  when the plasmoid is directly behind a Kerr black hole (at $\phi = 180^\circ$). 
Part of this image is significantly brightened, 
   as the corresponding emission is strongly Doppler boosted
   when the projected orbital velocity of the plasmoid 
   and the local rotational frame velocity, as viewed by the observer, are in alignment. 
In all cases of high viewing inclinations, 
  arcs and closed loops which are higher-order lensed images  
 which trace the boundary of the black hole shadow are present (cast by the black hole's photon capture sphere, not event horizon:
 see the bottom right panel in Figure \ref{fig-5}). 
A full set of Einstein rings are observable 
  when the plasmoid is at the azimuthal location $\phi \sim (120 - 150)^\circ$, 
  which is different to that in the case of the Schwarzschild black hole, 
  where the full set of Einstein rings occur when the plasmoid is located 
  directly behind the black hole (i.e.\ $\phi = 180^\circ$).

\begin{figure}
\begin{center} 
  \includegraphics[width=0.47\textwidth]{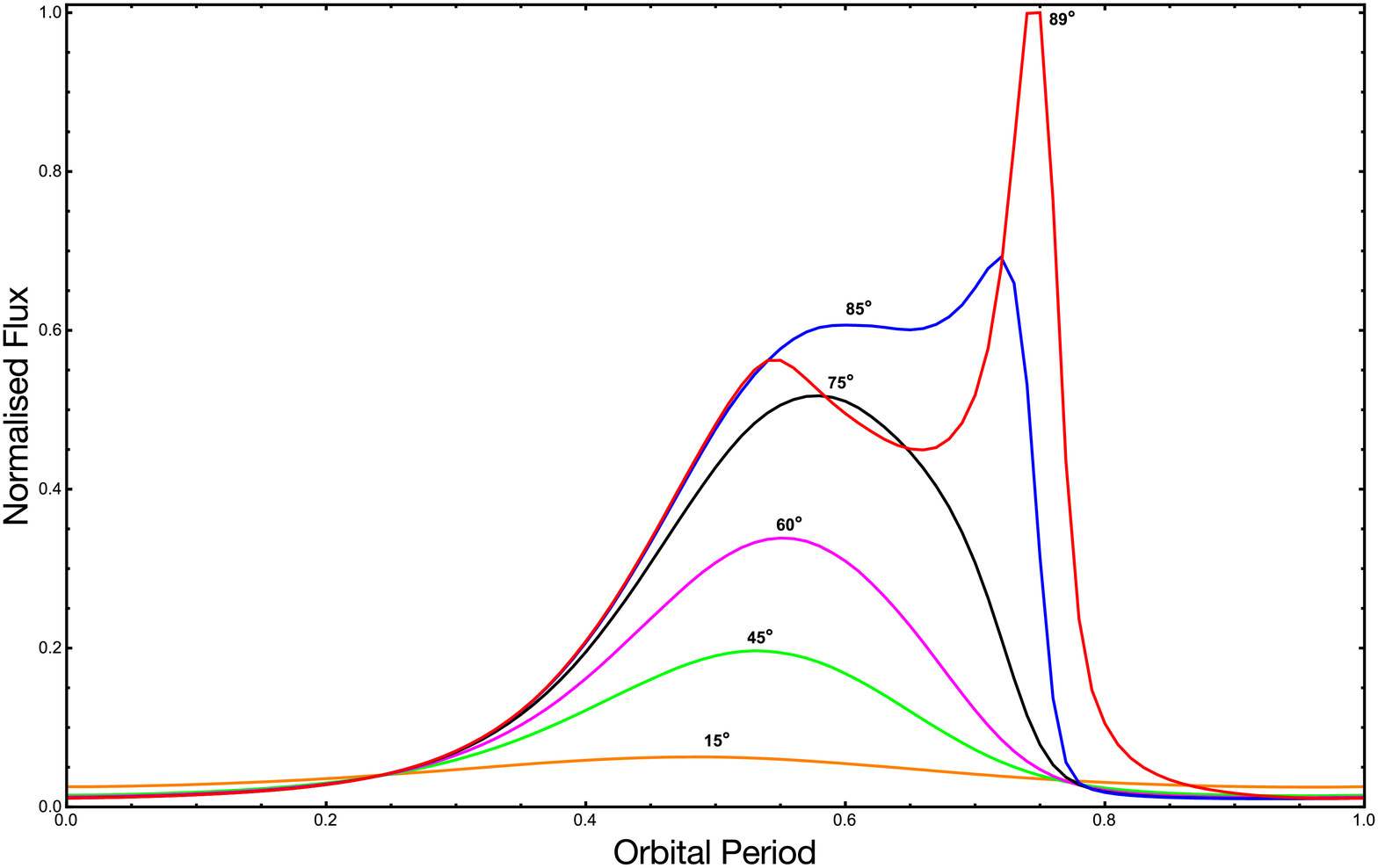}
  \vspace*{0mm}
  \includegraphics[width=0.47\textwidth]{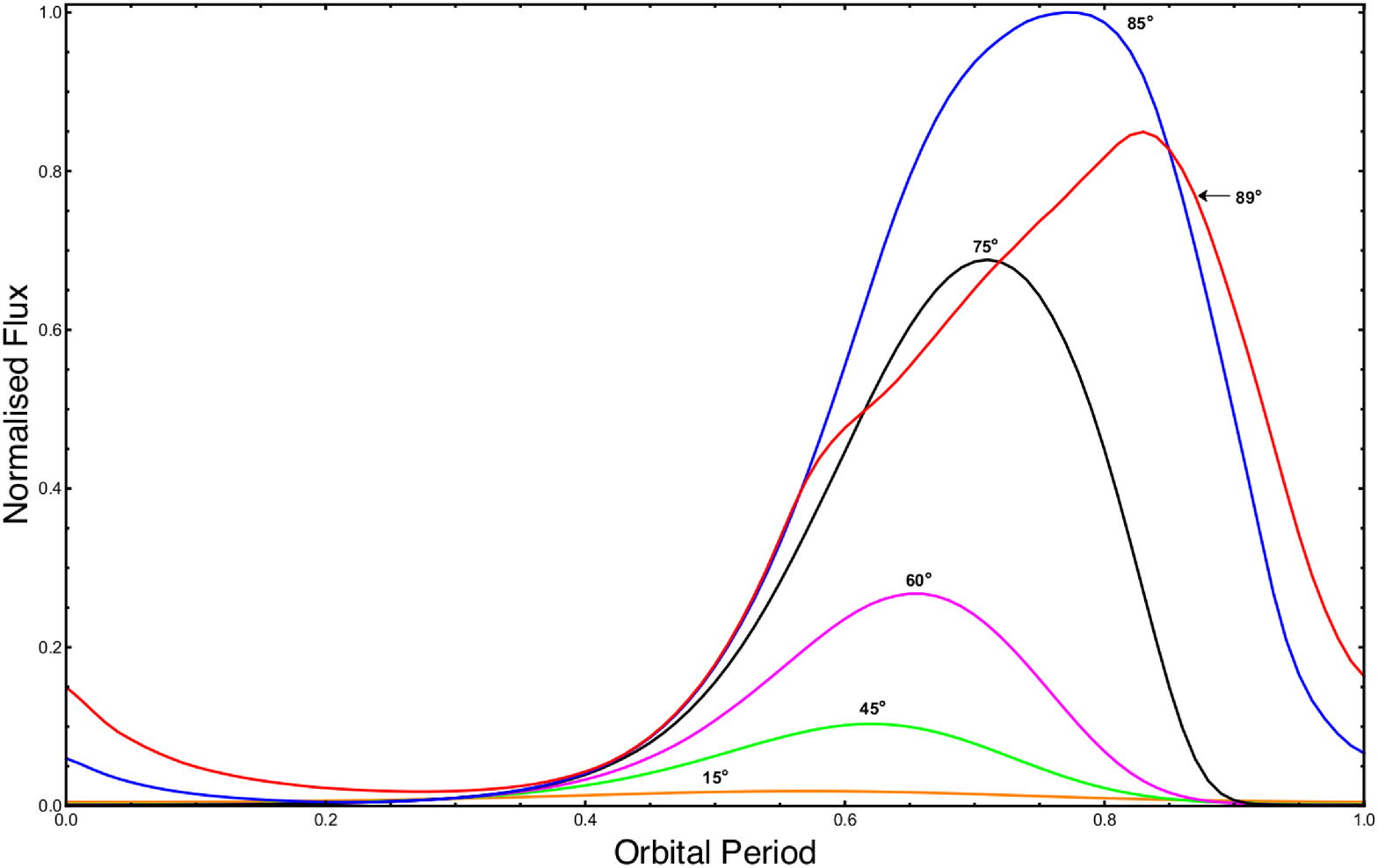}
\caption{Frequency-integrated lightcurves of a plasmoid orbiting a Schwarzschild black hole (top panel)
  and a Kerr black hole with $a=0.998$ (bottom panel) with arbitrary normalisation  
  for different viewing inclinations. 
The plasmoid is located at $r_{\rm c} = 6.5\,r_{\rm g}$ for the Schwarzschild black hole 
  and $2.5\,r_{\rm g}$ for the Kerr black hole, 
  which gives a full orbital period of $\approx 104\,r_{\rm g}/c$ 
  and of $\approx 31\,r_{\rm g}/c$ respectively.  
Phase 0.0 of the orbit corresponds to the plasmoid located at $\phi = 90^\circ$, 
  where a plasmoid orbiting the Schwarzschild black hole 
  would have the maximum receding velocity with respect to an observer. 
} 
\label{fig-6}
\end{center}
\end{figure}

\begin{figure}
\begin{center} 
  \includegraphics[width=0.47\textwidth]{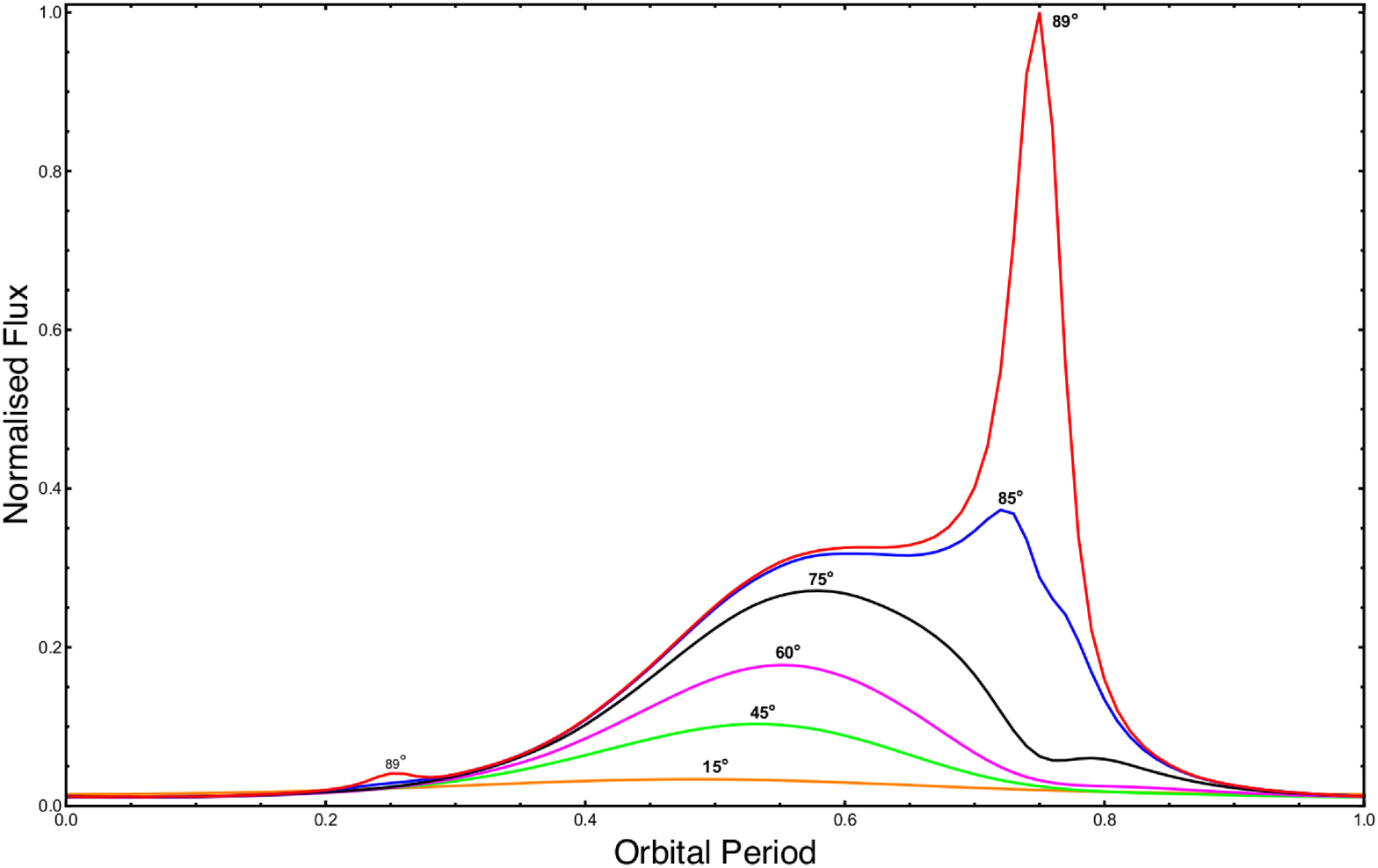}
  \vspace*{0mm}
  \includegraphics[width=0.47\textwidth]{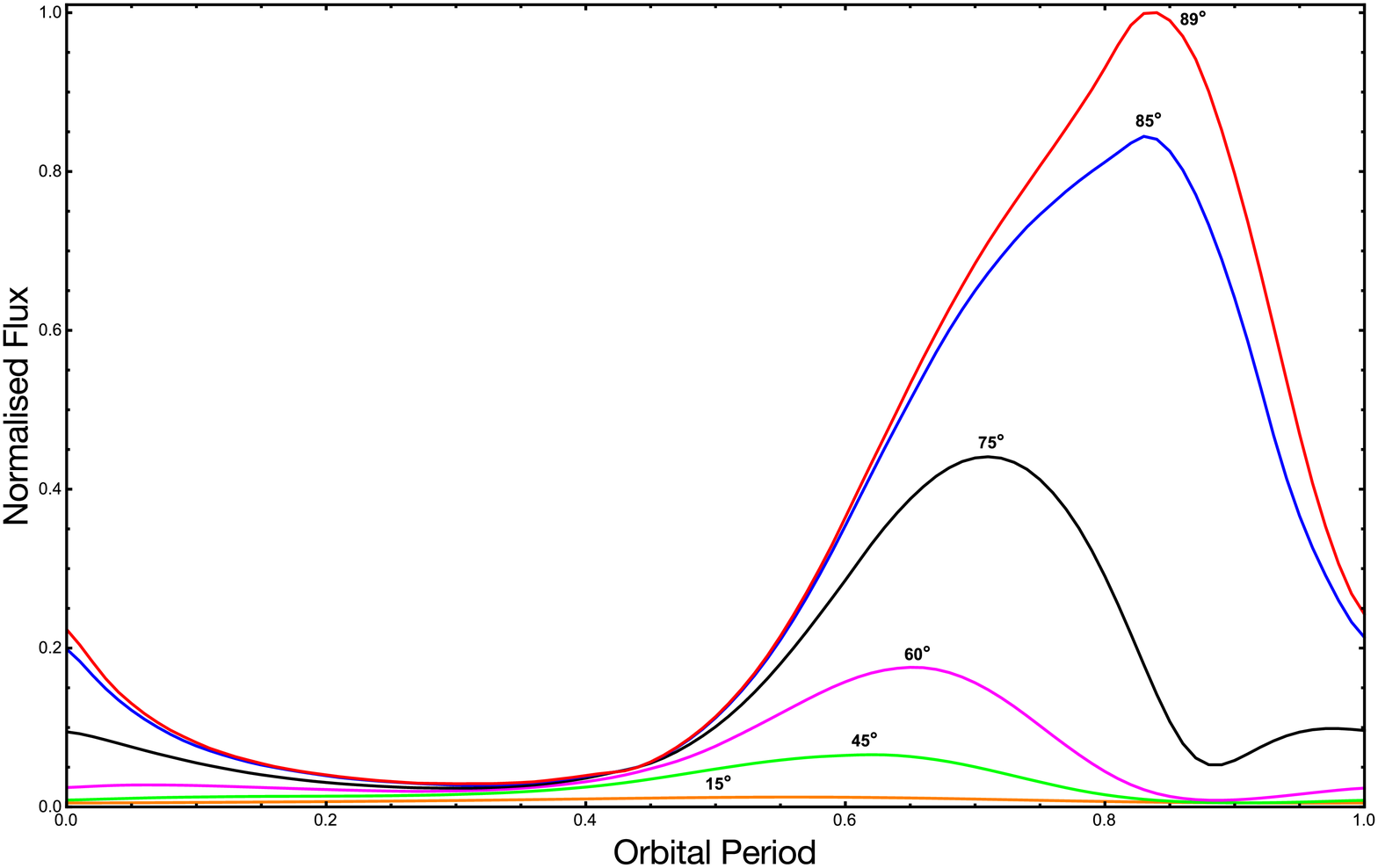}
\caption{Same as in Figure \ref{fig-6} but emission from all image orders are included. 
} 
\label{fig-7}
\end{center}
\end{figure}

The timescales on which the emission varies are determined by location of the plasmoid 
    with respect to the centre of the black hole. 
In this study, the plasmoid is placed at the ISCO, 
  and hence the orbital period is $\approx 104\, r_{\mathrm{g}}/c$ for the case of the Schwarzschild black hole 
  and $\approx 31\, r_{\mathrm{g}}/c$ for the case of the Kerr black hole with $a=0.998$. 
In addition to the periodic modulation, 
  the morphological structures in the lightcurves also provide
  information concerning the system geometry (in particular the viewing inclination) 
  and the black hole spin.

Figure \ref{fig-6}  shows 
  the frequency-integrated lightcurves of the emission from an opaque plasmoid viewed at different inclinations  
  constructed from the primary images 
  and Figure \ref{fig-7} the corresponding lightcurves 
  with the inclusion of the emission of the higher-order lensed images.  
The lightcurves have the following characteristics. 
(i) They do not peak exactly at the orbital phase 0.5.   
     The peak is skewed toward later orbital phases,   
       and the skewness is larger for higher viewing inclinations. 
(ii) The amplitude of the peak generally increases with viewing inclination  
       until it reaches $ i \approx 80^\circ - 85^\circ$,    
      and then, depending on the black hole spin, 
       the flux may saturate and decrease afterward when viewing inclination increases further. 
(iii) The lightcurves show more complex morphology at viewing inclinations close to $90^\circ$. 

What gives rise to such characteristics can be understood in terms of 
  relativistic beaming, Doppler boosting, transverse Doppler shift, 
  gravitational redshift, gravitational lensing and rotational frame dragging (in the case of the Kerr black hole).   
The projection of the orbital velocity into the line-of-sight varies with the orbital phase, 
   and this results in variations in the Doppler boosting of radiation intensity. 
It also regulates the amount of emission that is beamed in the direction of the plasmoid's motion 
  visible to the observer.   
Gravitational lensing brings 
 the Doppler boosted and beamed emission originating from the back side or the far side of the plasmoid,  
 which may otherwise be invisible, into the observer's line-of-sight.  
Gravitational redshift depends on the plasmoid's location 
  and is most prominent 
  for viewing inclinations close to $90^\circ$. 
 
In the case of the Schwarzschild black hole,  
  the receding velocity of the plasmoid with respect to the observer 
  is highest at phase 0 ($\phi = 90^\circ$). 
The suppression of the radiation intensity due to Doppler frequency redshift  
  is the strongest at this orbital phase,  
  and hence the observed flux is expected to be lowest.  
The increases in the observed flux between the phases 0 and 0.5 in the lightcurves 
  is evidence of an initial reduction in the intensity suppression 
  and a subsequent increase in the intensity boosted by relativistic Doppler shifts after phase 0.25. 
The peaking in the brightness 
  is caused by a combination of intensity boosting due to the relativistic Doppler effect 
  and the alignment of beamed emission into the line-of-sight.  
As the projected approaching velocity of the plasmoid to the line-of-sight reaches its maximum 
  at $\phi = 270^\circ$,  
  one may expect that the brightness peak would be located at the orbital phase of 0.5.  
However, the peaks are found to occur at a later orbital phase in all lightcurves.  
Also, the shifts of the brightness peak to a later orbital phase are larger for larger viewing inclination angles.  
If gravitational lensing is absent,   
  the observer sees only the direct emission from the front surface of the plasmoid facing the observer 
  when the plasmoid is located at $\phi \approx 0^\circ$, i.e.\ in front of the black hole.
  
Because of lensing the emission from the back and side surfaces of the plasmoid 
  can now reach the observer.  
The emission from the side surface is in the same direction as the plasmoid's motion 
   and is therefore strongly beamed and Doppler boosted. 
When it is visible it will increase the radiation intensity substantially.  
Although the first-order lensed emission and the direct emission are both primary emission, 
  the lensed emission, which has a long propagation path, 
  will take a long time to reach the observer.  
The shift is caused by the contribution of first-order lensed emission
  and the increase in the amount of shift with increasing viewing inclination angle
  by the increase in the contribution of the first-order lensed emission with increasing viewing inclination angle.  
For viewing inclinations close to $90^\circ$ 
  the lensed emission is so prominent that 
  it produces a second peak in the lightcurves.   

Note that the plasmoid's emission is also subject to time-dilation effects 
  (traverse Doppler shift and gravitational redshift)    
  which suppress the radiation intensity, 
  in addition to the special-relativistic effects associated 
  with the projection of the plasmoid's orbital velocity into the line-of-sight.  
Transverse Doppler shift, a special-relativistic effect, is relatively uniform 
  for low viewing inclination angles.  
It has larger variations when the plasmoid is viewed at high inclinations, 
   and it is strongest only when the plasmoid is directly in front of or behind the black hole. 
However, the effect is of second-order in the plasmoid velocity
  and is not expected to be a major contributor to the brightness variations in the lightcurve.   
Gravitational redshift is uniform throughout the orbital period,  
  as the plasmoid has a circular orbit and is a fixed distance from the black hole.  
Hence, it does not introduce brightness variations in the direct emission of the plasmoid.   
 
A key difference between the cases of the Kerr black hole and the Schwarzschild black hole 
   is that the plasmoid's orbit is closer to the event horizon in the case of the Kerr black hole.   
As a consequence, gravitational lensing and the intensity suppressions due to gravitational redshifts 
  are expected to be stronger. 
Another difference is that the Kerr black hole is rapidly rotating 
  and literally drags the space-time around it into rotation.  
On the one hand, rotational frame dragging 
  can redirect the emission, which would otherwise not reach the observer, 
  into the observer's light-of-sight, 
  but on the other hand it can also spatially smear the beamed emission from the plasmoid.  

The lightcurves of the plasmoid orbiting a Kerr black hole  
   are similar to those of the plasmoid orbiting around a Schwarzschild black hole 
   for viewing inclinations $i \approx  60^\circ$ or lower  
   (see the top and bottom panels of Figure \ref{fig-6}). 
This suggests that 
   at these viewing inclinations     
   the brightness modulations are determined mainly by 
   Doppler boosts and relativistic beaming associated with the plasmoid's orbital motion.   
At high viewing inclination angles ($i \sim 80^\circ - 90^\circ$) 
  the differences between the lightcurves in the two cases become more noticeable. 
Firstly, there is only one peak present in the lightcurves of the plasmoid orbiting a Kerr black hole.   
Secondly, the peaks broaden and  
  there are also substantial amounts of emission extending to orbital phases of 0 and beyond.   
Interestingly, the peaks are relatively symmetrical and have a round top 
  at moderate viewing inclination angles ($i \sim 60^\circ - 75^\circ$), 
  but begin to skew toward the later orbital phases as the viewing inclination increases further. 
The peak amplitude also drops when $i$ approaches $90^\circ$. 

Note that gravitational lensing is important at the highest viewing inclinations.  
For the Kerr black hole,  
  rotational frame dragging effects are prominent in space-time 
  near the equatorial plane. 
At high viewing inclination angles, 
  photons that can reach the observer
  follow paths which are close to the equatorial plane 
  and the emission is thus affected by rotational frame dragging 
  more strongly at high viewing inclinations than at low viewing inclinations.  
This distorts the visual appearance of the plasmoid in the image 
  captured by the observer at a specific instance 
  in the local reference frame of the observer.  
Beam smearing by differential rotational frame dragging 
  explains the substantial drift of the peak in the lightcurve to the later orbital phases.  
Strong gravitational redshifts and rotational frame dragging  
  counteract relativistic beaming and Doppler boosting of the emission 
  at the phases $\sim 0.55 - 0.85$ 
  at viewing inclinations close to $90^\circ$. 
This causes the peak to appear asymmetric at these viewing inclinations. 
The absence of the second peak in the lightcurves at the viewing inclinations close to $i = 90^\circ$ 
  is due to the fact that the emission that is beamed in the forward direction of the plasmoid's motion 
  is not strongly focused by gravitational lensing.  
It also reduces the amplitude of the peak at the viewing inclination of $89^\circ$,  
 making it smaller those at the viewing inclinations of about $85^\circ$. 

The inclusion of the emission in the higher-order images  
  is expected to increase the peak amplitude of the lightcurve. 
However, the increase is only substantial at the high viewing inclinations 
   where lensing effects are important.  
The contribution of the higher-order images is more visible 
  when comparing the location of the peaks in the lightcurves 
  in Figures \ref{fig-6} and \ref{fig-7}. 
The peaks occur in later phases when the emission in the high-order images is included. 
This can be understood as follows. 
The higher-order lensed emission has a larger propagation path length 
  and hence takes longer to reach the observer. 
Note that there are also additional peaks that can be formed due to high-order lensing. 
At $i=75^{\circ}$ a secondary peak can be seen at the late phases in the lightcurve   
   for both Schwarzschild and Kerr black holes. 
At $i = 89^\circ$ 
   there is an extra small peak at around phase 0.25 in the lightcurves 
   for the plasmoid orbiting a Schwarzschild black hole.  
The effects of multiple-order lensing of high-order lensed emission 
  are more prominent in the case of the Kerr black hole than in the case of the Schwarzschild black hole 
  as the lensed emission is also subject to rotational frame dragging.  
In some situations, 
  radiation emitted by the plasmoid earlier in an orbit   
  can be severely lensed,  
  such that it may arrive at around the same time 
  as the direct radiation that is emitted by the plasmoid later along its orbital path.

\subsection{Optically thin plasmoid}

The entire volume of an optically thin plasmoid 
 contributes to the radiation,  
  whereas for an opaque emitter only its surface contributes. 
In the vicinity of a black hole, 
  the differential weighting and geometrical projection of the emission 
  from within an optically thin plasmoid 
  will couple with the plasmoid's relativistic orbital motion  
  and the black hole's gravitational effects 
  (in particular gravitational lensing and rotational frame dragging) 
  to shape the plasmoid's emission lightcurve. 
However, such effects are obvious only for sufficiently high viewing inclinations. 

\begin{figure}
\begin{center} 
  \includegraphics[width=0.47\textwidth]{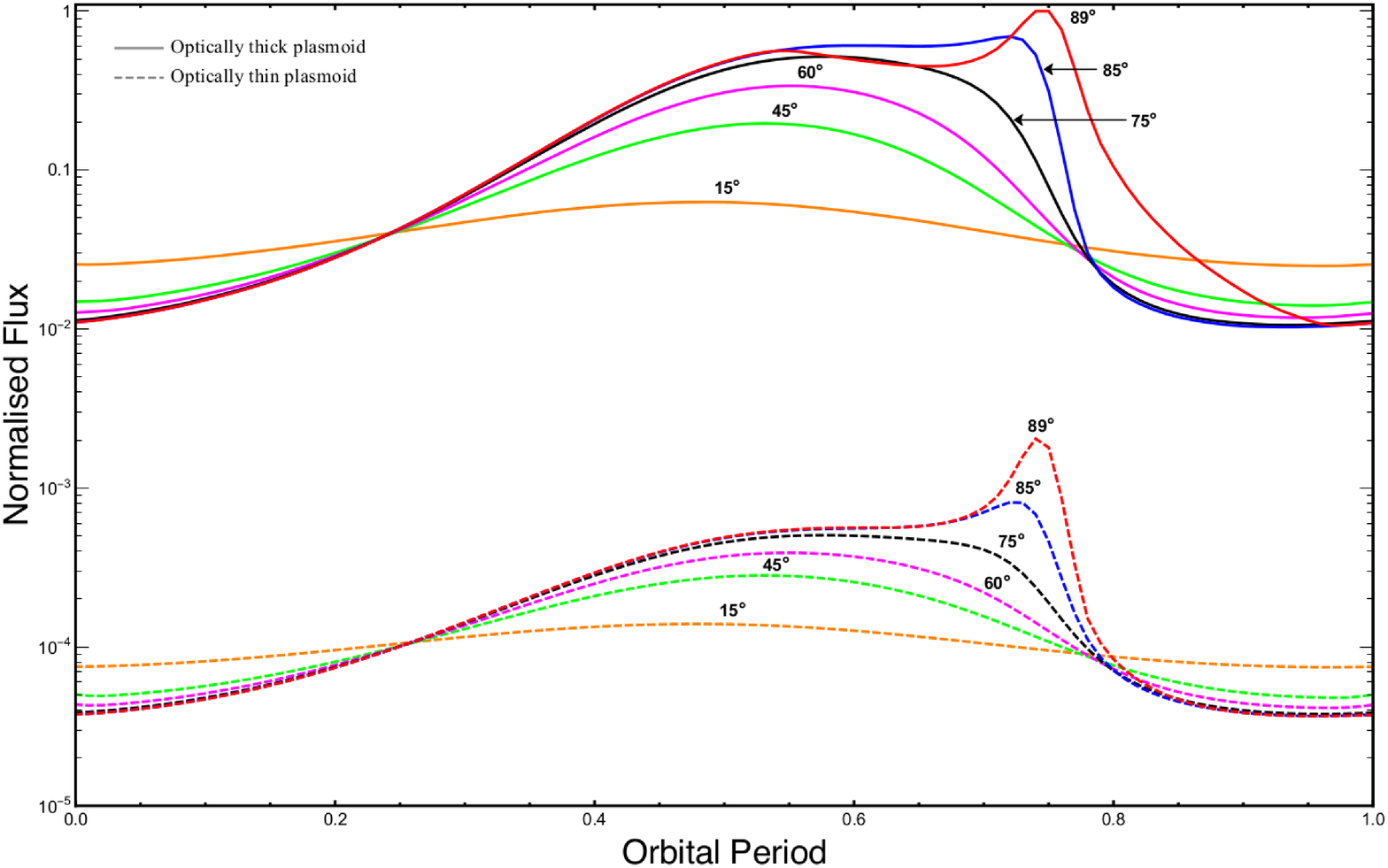}
  \vspace*{0mm}
  \includegraphics[width=0.47\textwidth]{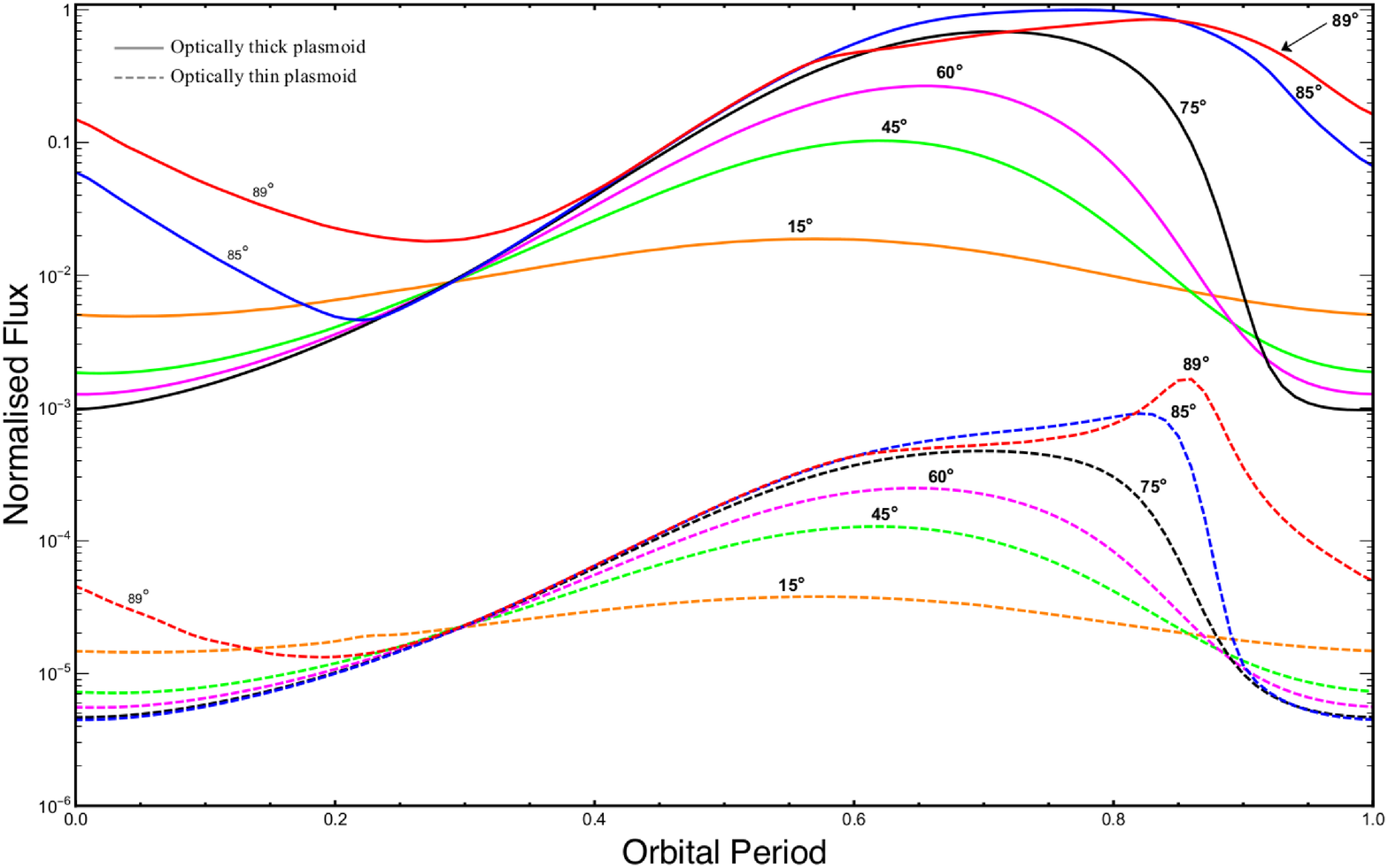}
\caption{Lightcurves of emission of an optically thin plasmoid (dashed lines) 
  orbiting a Schwarzschild black hole (top panel) 
  and a Kerr black hole with $a = 0.998$ (bottom panel) 
  for different viewing inclinations. 
Only the primary image is used in the construction of each lightcurve.  
The flux is logarithmically scaled. 
Colour coding for the lightcurves is the same as in Figures \ref{fig-6} and \ref{fig-7}.  
The corresponding lightcurves of an opaque plasmoid (see also Figure \ref{fig-6}) are shown for comparison (solid lines).
} 
\label{fig-8}
\end{center}
\end{figure}

Figure \ref{fig-8} shows that at viewing inclination angles below about $75^\circ$,  
  the lightcurves of an optically thin plasmoid and an opaque plasmoid are similar 
  regardless of the spin of the black hole. 
At low viewing inclinations, gravitational lensing effects are relatively unimportant. 
Even when the entire volume of the optically thin plasmoid emits, 
  not much strongly beamed and Doppler boosted emission is additionally lensed into view 
  in the primary image.   
Gravitational lensing effects are stronger at the higher viewing inclinations.  
At viewing inclinations close to $i = 90^\circ$,  
  the difference between the lightcurves of plasmoids with different optical thickness becomes noticeable  
  for Schwarzschild and Kerr black holes. 

For an isotropic emissivity, 
  the emission from interior of the plasmoid, 
  which are relativistic beamed and Doppler boosted,  
  can be lensed into view
  without being restricted by the available projected emission surface that is ``visible'' to the observer. 
The emission from the optically thin plasmoid is therefore more strongly peaked than that of the opaque plasmoid. 
This leads to a more rapid drop in the plasmoid's apparent brightness in the primary image after the brightness peak  
  as compared to the opaque plasmoid.

\begin{figure}
\begin{center} 
  \includegraphics[width=0.47\textwidth]{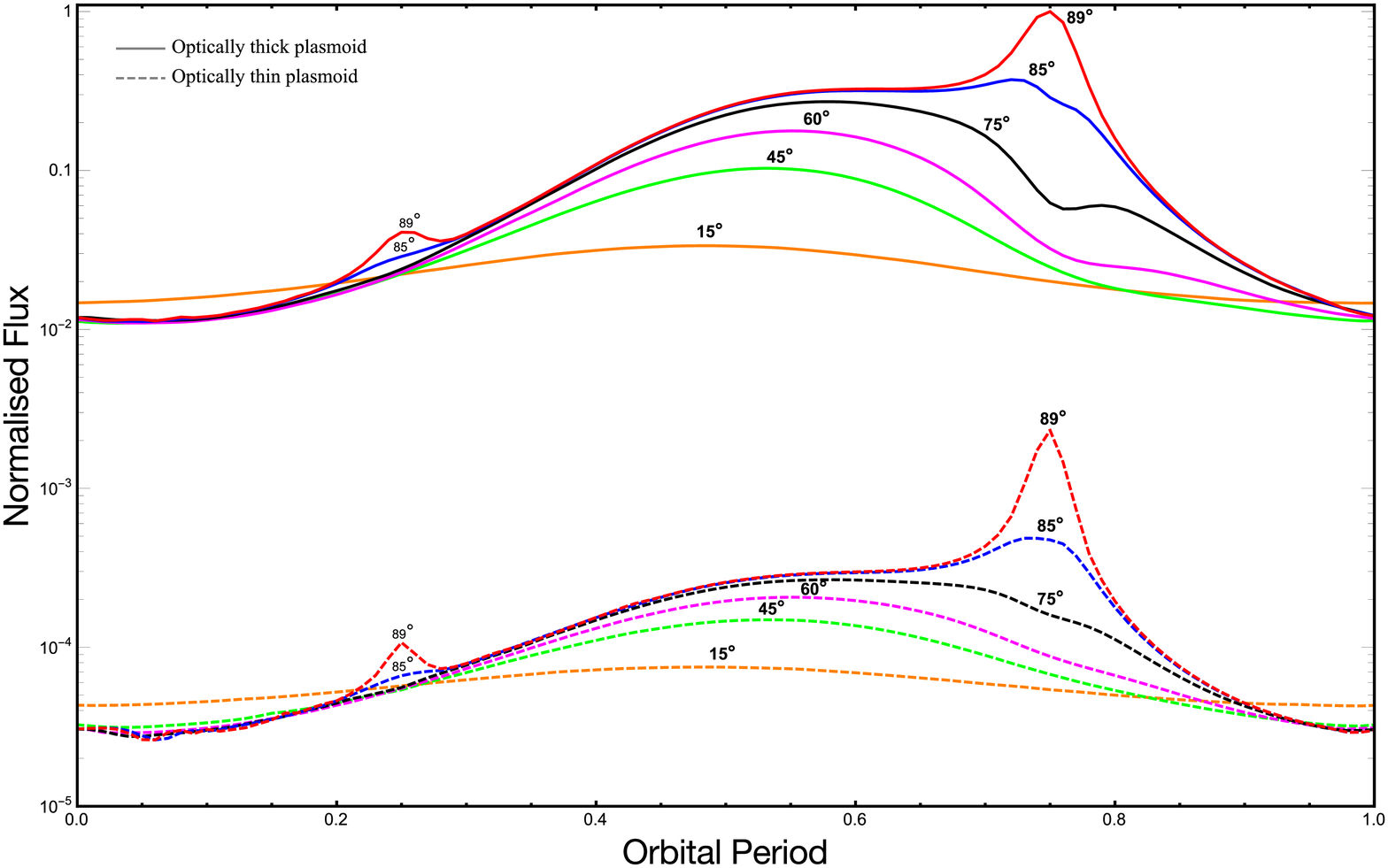}
  \vspace*{0mm}
  \includegraphics[width=0.47\textwidth]{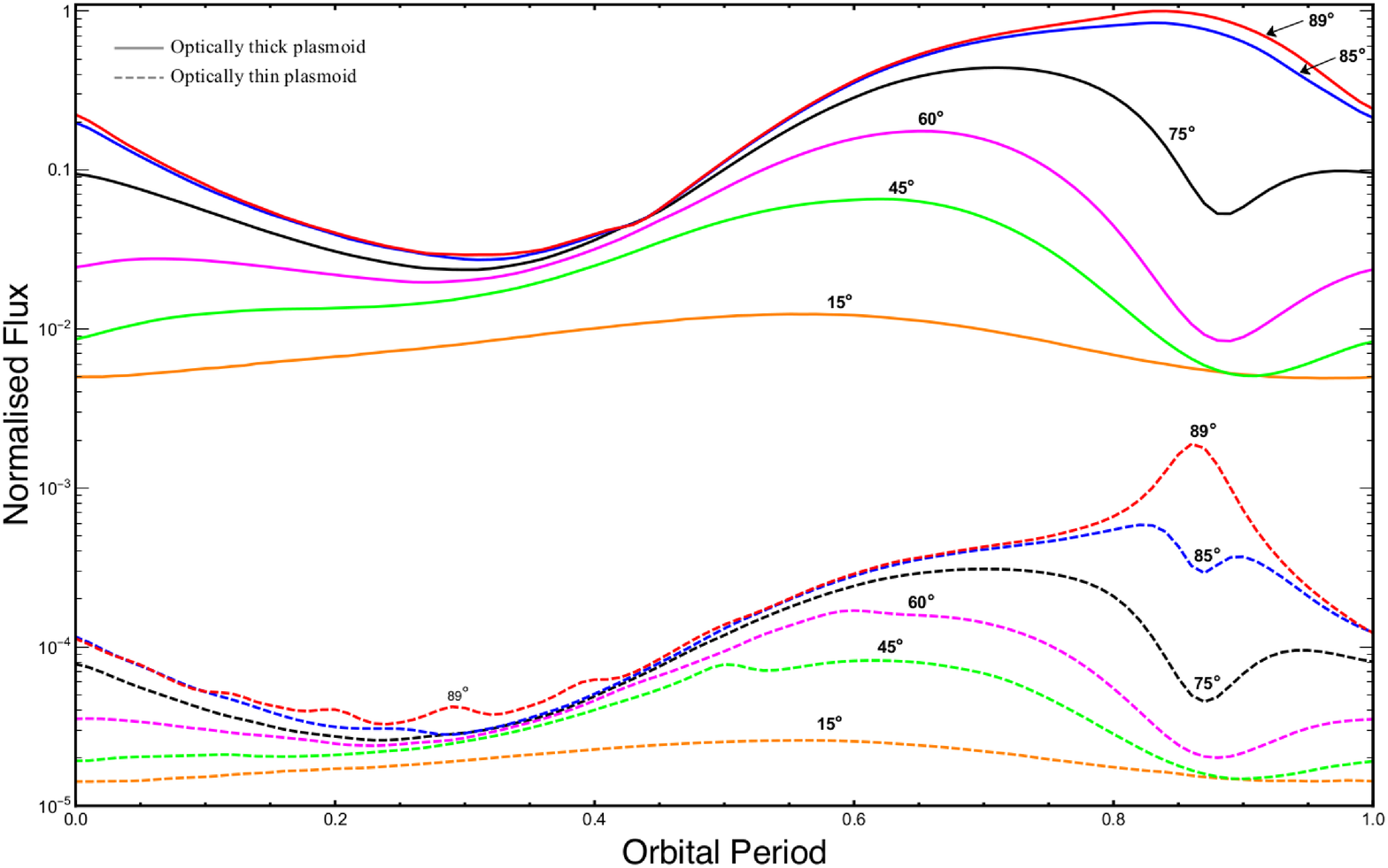}
\caption{As in Figure \ref{fig-8}, but now including the contribution from all image orders. } 
\label{fig-9}
\end{center}
\end{figure}

With the inclusion of the emission in the higher-order images (Figure \ref{fig-9}), 
   the differences between the lightcurves of the optically thin and the opaque plasmoids 
   are more obvious when the plasmoid is transiting the region of maximum blueshift into full view in front of the observer 
   (i.e.\ at phases $0.5  - 1.0$).   
The multiple small peaks observed in the lightcurves of the optically thin plasmoid orbiting a Kerr black hole 
  correspond to accumulation of emission received at later times 
  with respect to the direct emission from the plasmoid. 
This effect becomes more pronounced as the observer's inclination angle increases 
  as the contributions of high order lensed images increase. 
  
\begin{figure}
\begin{center} 
  \includegraphics[width=0.47\textwidth]{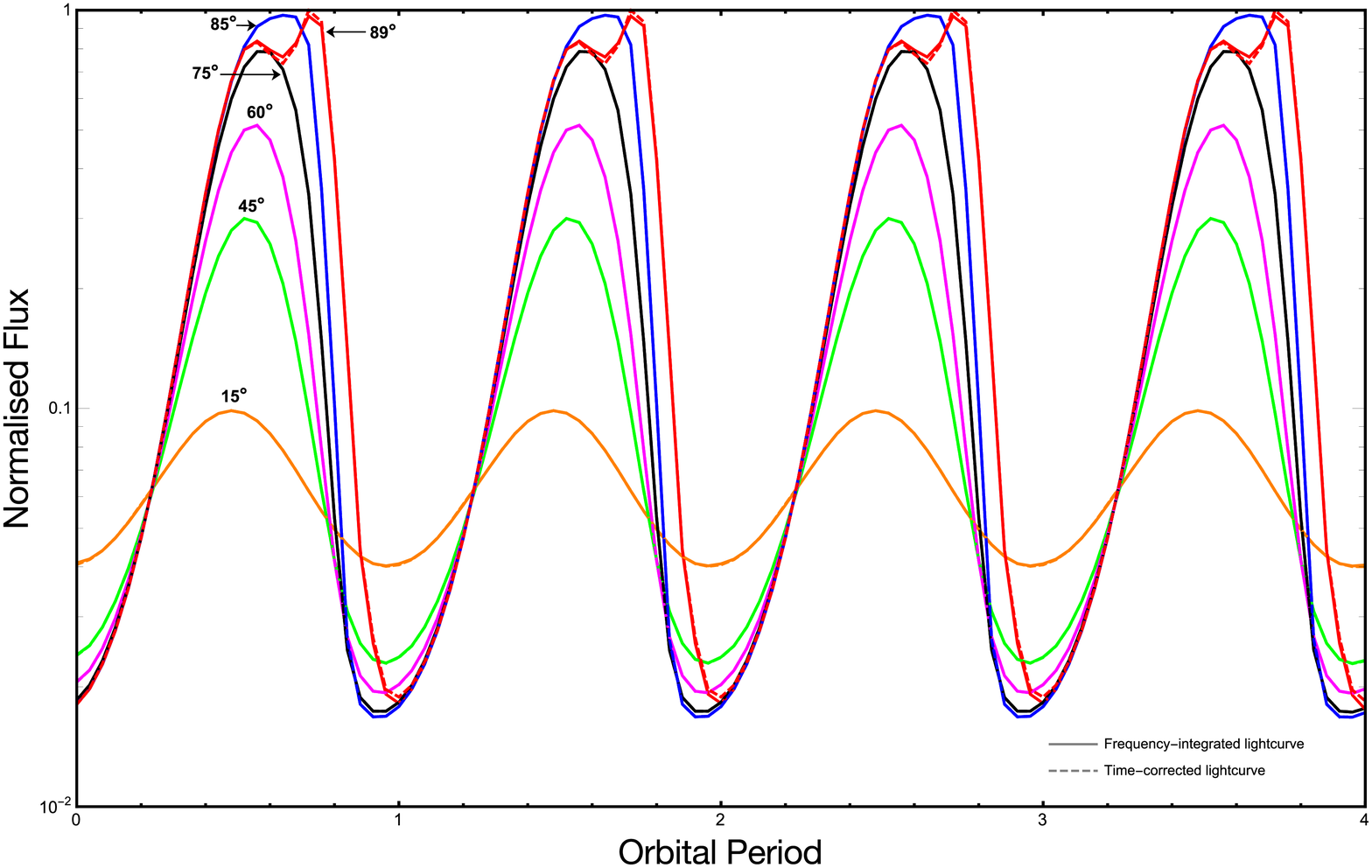}
  \vspace*{0mm}
  \includegraphics[width=0.47\textwidth]{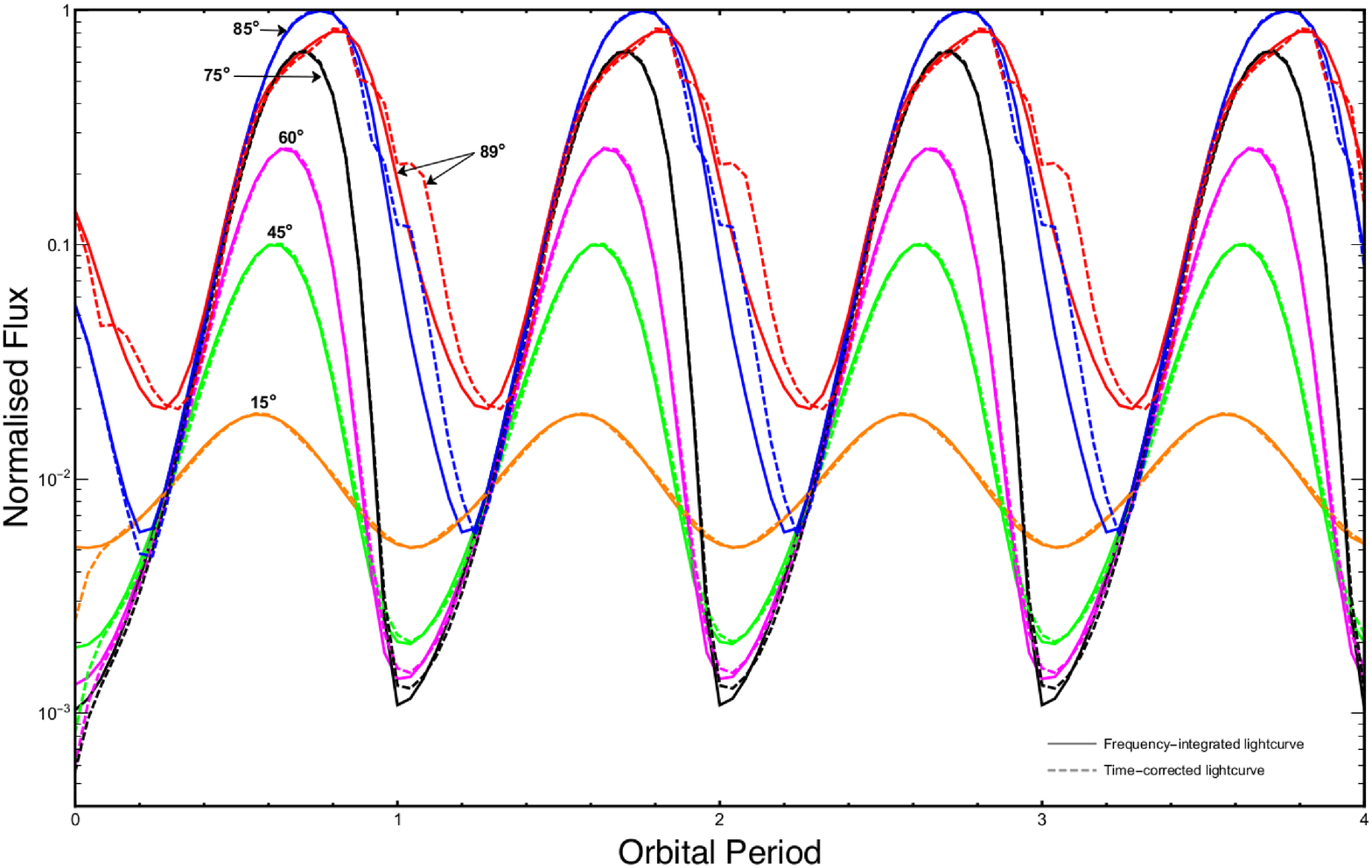}
\caption{Comparison of frequency-integrated lightcurves (solid lines) and time-corrected lightcurves (dashed lines) of an optically thick plasmoid
viewed over four orbital periods of the plasmoid. 
Only direct emission is considered. 
Top and bottom panels (for a Schwarzschild and a Kerr black hole respectively) 
  take the same parameters as previous lightcurve calculations. 
Same colour-coding and panel order as the lightcurves in previous Figures.
} 
\label{fig-10}
\end{center}
\end{figure} 

\begin{figure}
\begin{center} 
  \includegraphics[width=0.47\textwidth]{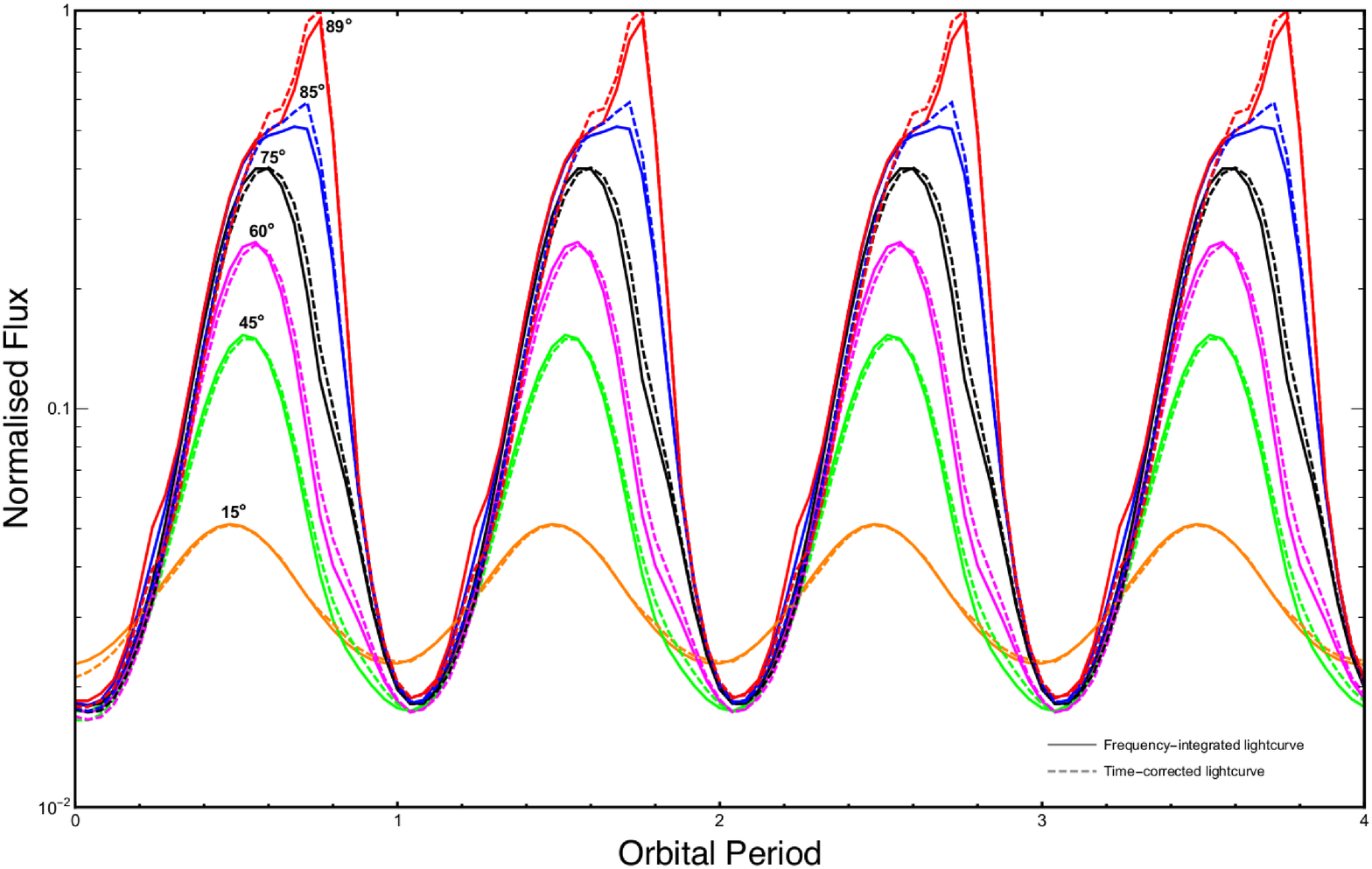}
  \vspace*{0mm}
  \includegraphics[width=0.47\textwidth]{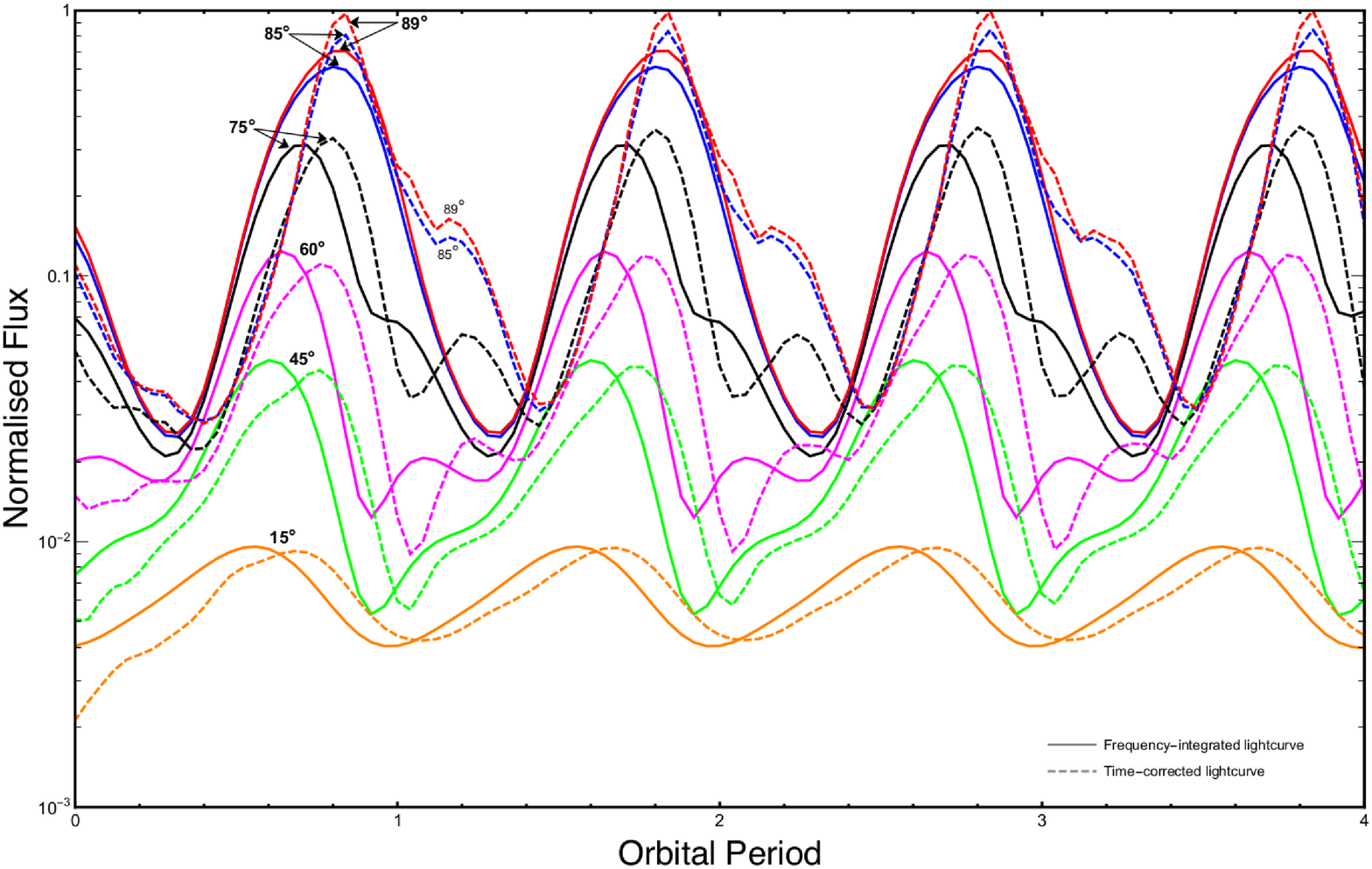}
\caption{As in Figure \ref{fig-10}. All image orders are now considered.} 
\label{fig-11}
\end{center}
\end{figure}

\subsection{Time-corrected lightcurves}
\label{sec-TCLC}
 
Radiative emissions from different plasmoid locations have different path lengths 
  to travel in order to reach the observer, 
  and these paths are also bent by the black hole's gravity.
Differences in these path lengths lead to differences in light travel times. 
Thus, at any instant in the observer's reference frame, 
  two photons arriving simultaneously could have originated 
  from different parts of the plasmoid at different locations in the plasmoid's orbit.    
This differential arrival time of the photons would modify the observed lightcurve of the plasmoid. 
The lightcurves presented so far have not taken into account this effect. 
It is important to quantify how the different photon arrival times 
  will affect the observed emission from the plasmoid during its motion, 
  as well as to determine to what extent a proper time-corrected frequency-integrated lightcurve 
  will differ from an ``uncorrected" lightcurve. 
  It is also important to determine which system parameters this difference depends on most strongly 
  if we wish to extract system information from the lightcurves obtained in observations.  

In Figure \ref{fig-10} 
  we compare the ``uncorrected" (solid lines) 
  and ``corrected" (dashed lines) lightcurves from an optically thick plasmoid,    
  considering direct emission only.
For the Schwarzschild case (top panel) the difference is very small 
  and is most prominent at the peak of the emission at $i=89^{\circ}$. 
For the Kerr case (bottom panel) the effect is more noticeable but still mild.
The differences are again most pronounced at the highest inclination angles 
  and can be interpreted as a small but non-negligible phase shift of the lightcurve to later times.
This is due to the frame-dragging effect of the Kerr black hole, 
  elongating the geodesics each photon follows before reaching the observer and thus delaying their arrival.
As the observer's inclination angle increases, 
  a larger fraction of all received photons experience the effects of frame-dragging 
  before reaching the observer 
  and so this effect of photon time-delay on the lightcurves is more pronounced.

Emission from all image orders is accounted for in the lightcurve comparisons presented in Figure \ref{fig-11}. 
For the case of a Schwarzschild black hole 
   the difference between the lightcurves is very small and amounts to a small shift in phase of the lightcurve, 
   with a modest boost in the emission peak seen 
   at high viewing inclination angles. 
For the case of the Kerr black hole 
   the difference between the two lightcurves is substantial 
   and represents a phase difference on the order of $20\%$ of the orbital period of the plasmoid. 
This also leads to a more notable shift in the peak of the lightcurves at high observer inclination angles. 
We may conclude that  
  when observing plasmoids orbiting close to the event horizon of rapidly rotating black holes, 
  general-relativistic radiative transfer calculations must properly account for the photon arrival time. 
However, in the case of slowly rotating black holes and plasmoids not in close proximity to the event horizon, 
  a conventional frequency-integrated lightcurve without consideration of relative photon arrival times 
  is an acceptable approximation.

\section{Emission from an ejected plasmoid} 

To avoid complexities that mask the relativistic and dynamical effects 
  that occur at different stages when the plasmoids are launched 
  and the comparison between the ejected plasmoids and the orbiting plasmoids,    
  we adopt a simple scenario 
  where the internal properties of the ejected plasmoids do not evolve substantially 
  over the dynamical timescale of the system. 
In other words, the structure and emissivities of the plasmoids 
  remain unchanged before and after launch, 
  and the ejected plasmoids are practically identical to the orbiting plasmoid 
  considered previously. 
As in \S 3 we consider two demonstrative cases: 
  an optically thick plasmoid and an optically thin plasmoid, 
  both subject to the gravity of a Schwarzschild black hole and a Kerr black hole (with $a=0.998$). 
The ejected plasmoids now have a vertical motion, 
  due to magneto-hydrodynamical processes as 
  described in \S 2.2.3. 
The radiative transfer calculations and the procedures 
  to construct the lightcurves is the same as that for the orbiting plasmoids.

Note that the motion of a magnetically ejected plasmoid 
   had been investigated for a particular set of parameters in 
   the context of infrared flaring emission \citep[see][]{Vincent2014}.  
The plasmoid in the adopted model of the aforementioned study 
  was ejected and then fell back onto the black hole
  after attaining a maximum height.
Here, with the sporadic outflows observed in some black hole systems in mind, 
  we consider those plasmoids that subsequently break the confinement of the black hole's gravity 
  and escape from the system.   

As shown in the previous sections, gravitational lensing and the arrival time differences of plasmoids 
   with finite sizes are dependent on the location of the plasmoids 
   with respect to the viewing geometry of the black hole and the observer.  
For the ejected plasmoids, an addition parameter (i.e.\ the azimuthal parameter $\phi$) is needed 
  to specify the launching site and the relative phases in the lightcurves.  
As a demonstration, we show calculations for two cases, 
  corresponding to where the reconnection and subsequent ejection of the plasmoid occurs, 
  namely at the positions $\phi=90^{\circ}$ and $\phi=270^{\circ}$. 
For the Schwarzschild black hole,  
  $\phi=90^{\circ}$ corresponds to the point of maximum redshift 
  and $\phi=270^{\circ}$ to the point of maximum blueshift. 
These cases provide a good contrast in the emission at the ejection, at least for the case of Schwarzschild black hole. 
For simplicity, we consider the same cases (i.e.\ $\phi=90^{\circ}$ and $270^{\circ}$ in the case of Kerr black hole). 
The plasmoid's initial conditions are taken to be identical to those of the systems in \S 3. 
Emission of all image orders are included in the construction of the frequency-integrated lightcurves  
  of the plasmoids.  

\subsection{Ejected optically thick plasmoid}

\begin{figure}
\begin{center} 
  \includegraphics[width=0.47\textwidth]{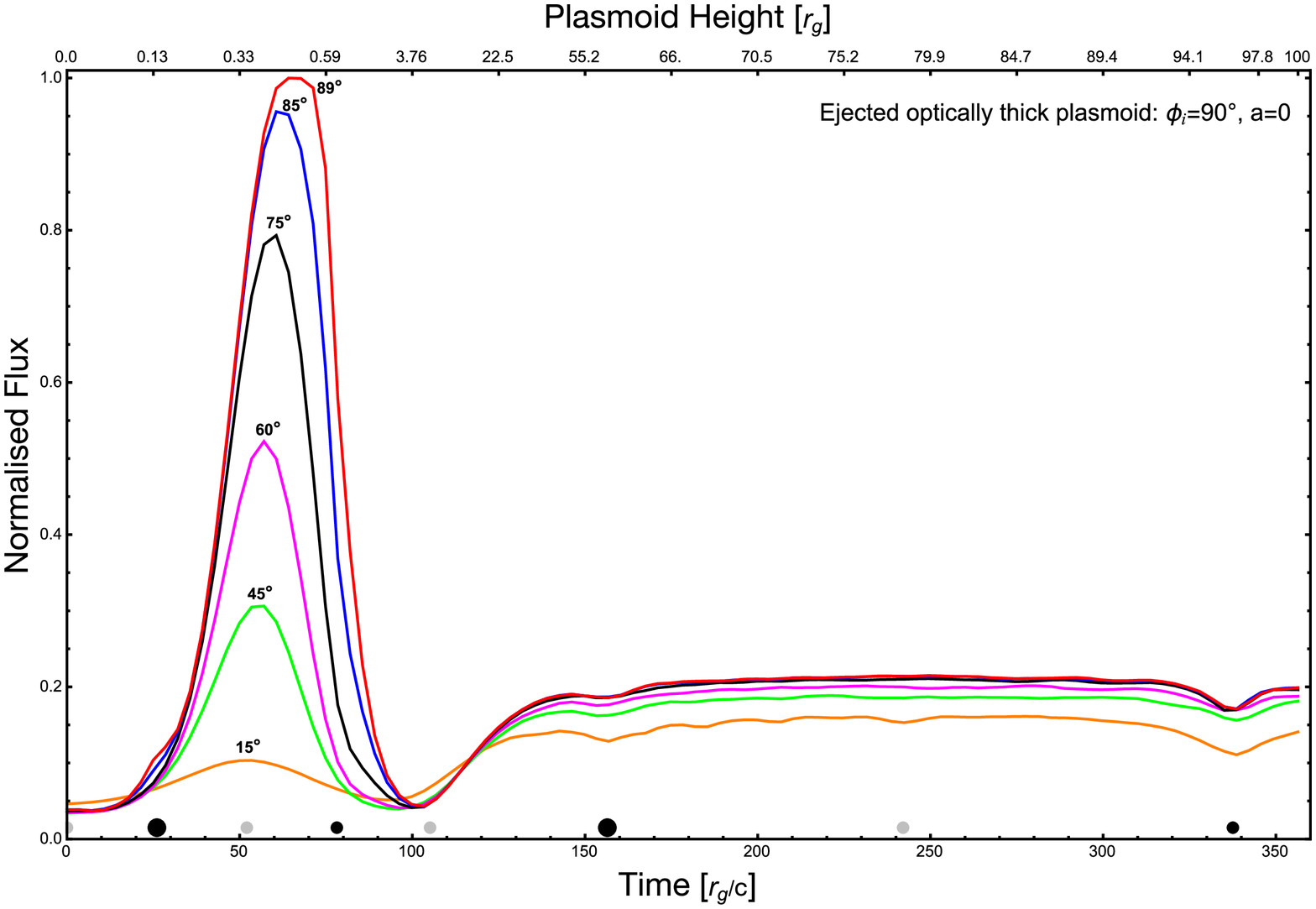}\vspace{1mm}
  \vspace*{0mm}
  \includegraphics[width=0.47\textwidth]{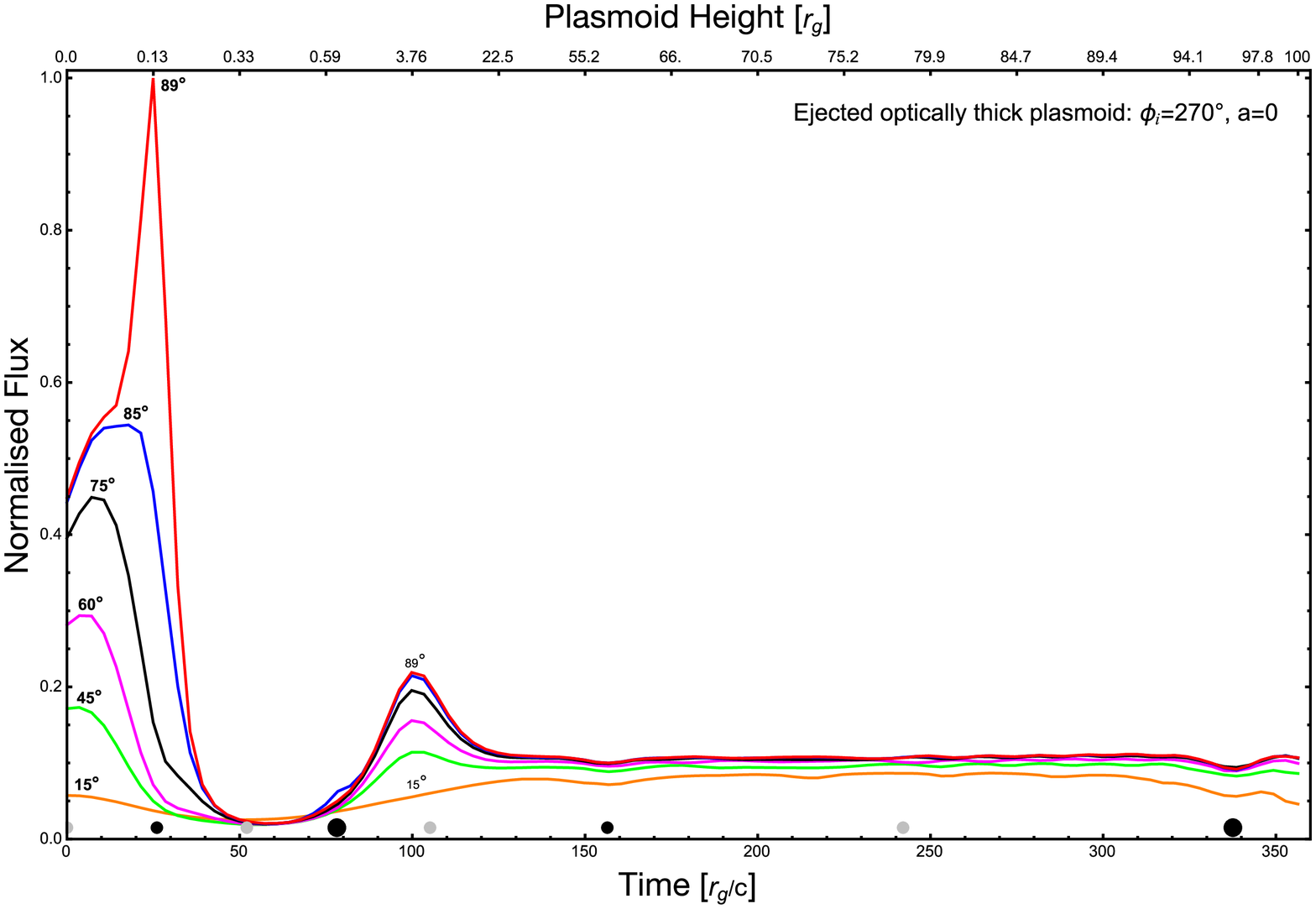}
\caption{Frequency-integrated lightcurves from an optically thick plasmoid ejected from a Schwarzschild black hole starting from an initial position of $\phi_{i}=90^{\circ}$ (top panel) 
and a plasmoid initial position of $\phi_{i}=270^{\circ}$ (bottom panel). All image orders are considered. Plasmoid height as a function of time is shown on the upper horizontal axis.
The small black circles denote when the plasmoid is located at $\phi=0^{\circ}$. 
Larger black circles correspond to a plasmoid location of $\phi=180^{\circ}$. The smaller grey circles located between the aforementioned small and larger black circles correspond accordingly to plasmoid positions of $\phi=90^{\circ}$ and $\phi=270^{\circ}$.} 
\label{fig-12}
\end{center}
\end{figure}

Figure \ref{fig-12} shows the lightcurve from an optically thick plasmoid ejected from a Schwarzschild black hole.
After the rapid acceleration phase between $\sim100\, r_{\mathrm{g}}/c$ and $150\, r_{\mathrm{g}}/c$ 
  the lightcurves become flat and almost featureless, 
  except for small dips in flux when the plasmoid passes 
  in front of ($\phi=0^{\circ}$) and behind ($\phi=180^{\circ}$) the black hole.
These small dips, which are present in all ejected plasmoid lightcurves, 
   are due to the transverse Doppler shift of the emission as the plasmoid passes across the observer's line of sight. 
The relative amplitude of the fluxes between the lightcurve peak prior to rapid acceleration 
  and the (relatively) flat phase of the lightcurve after this acceleration phase 
  are markedly different in the top and bottom panels of Figure \ref{fig-12}. 
As this depends on $\phi$, 
  it is a marker of where the plasmoid is launched 
  (presumably caused by the magnetic reconnection event at the location).

\begin{figure}
\begin{center} 
  \includegraphics[width=0.47\textwidth]{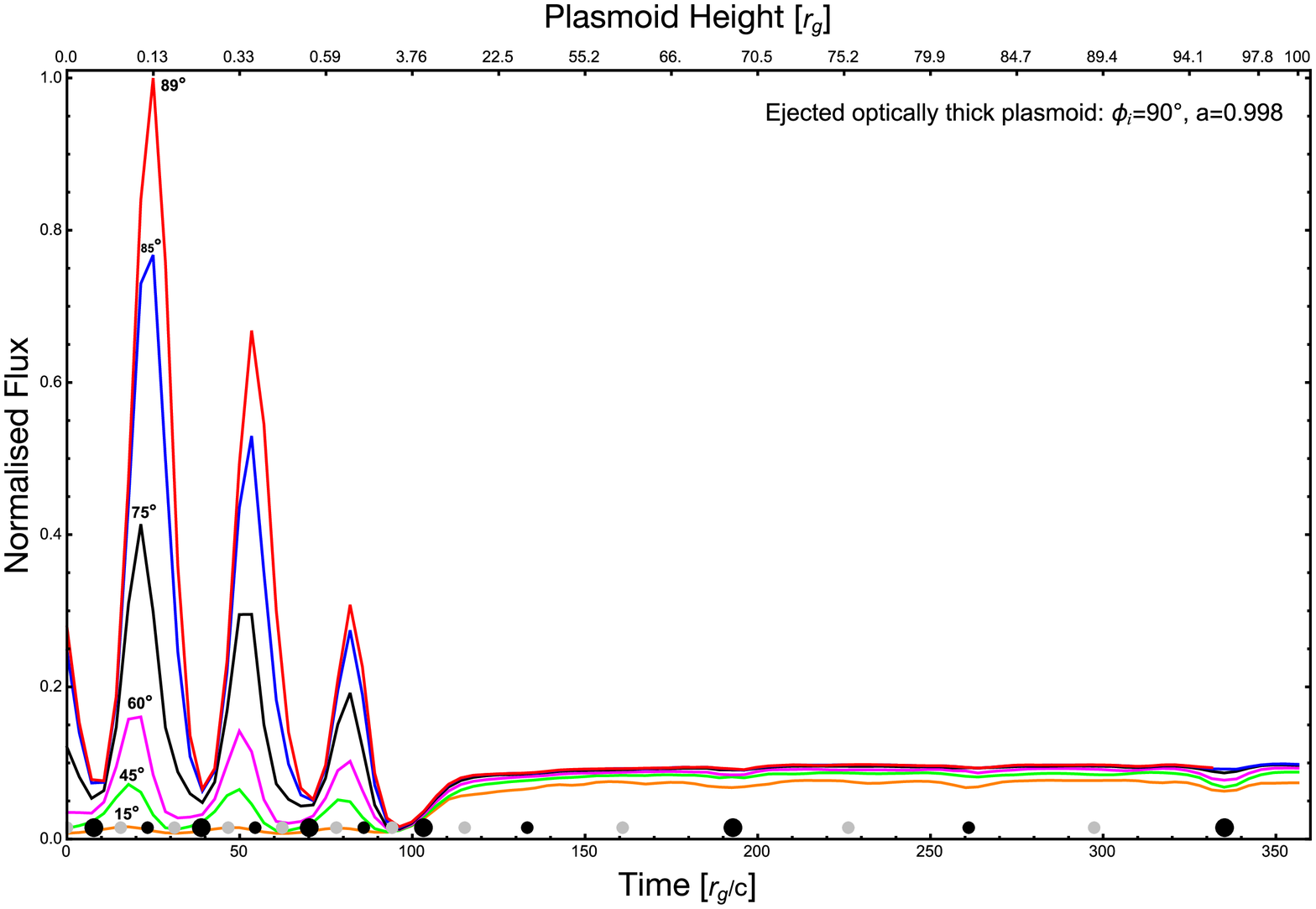}\vspace{1mm}
  \vspace*{0mm}
  \includegraphics[width=0.47\textwidth]{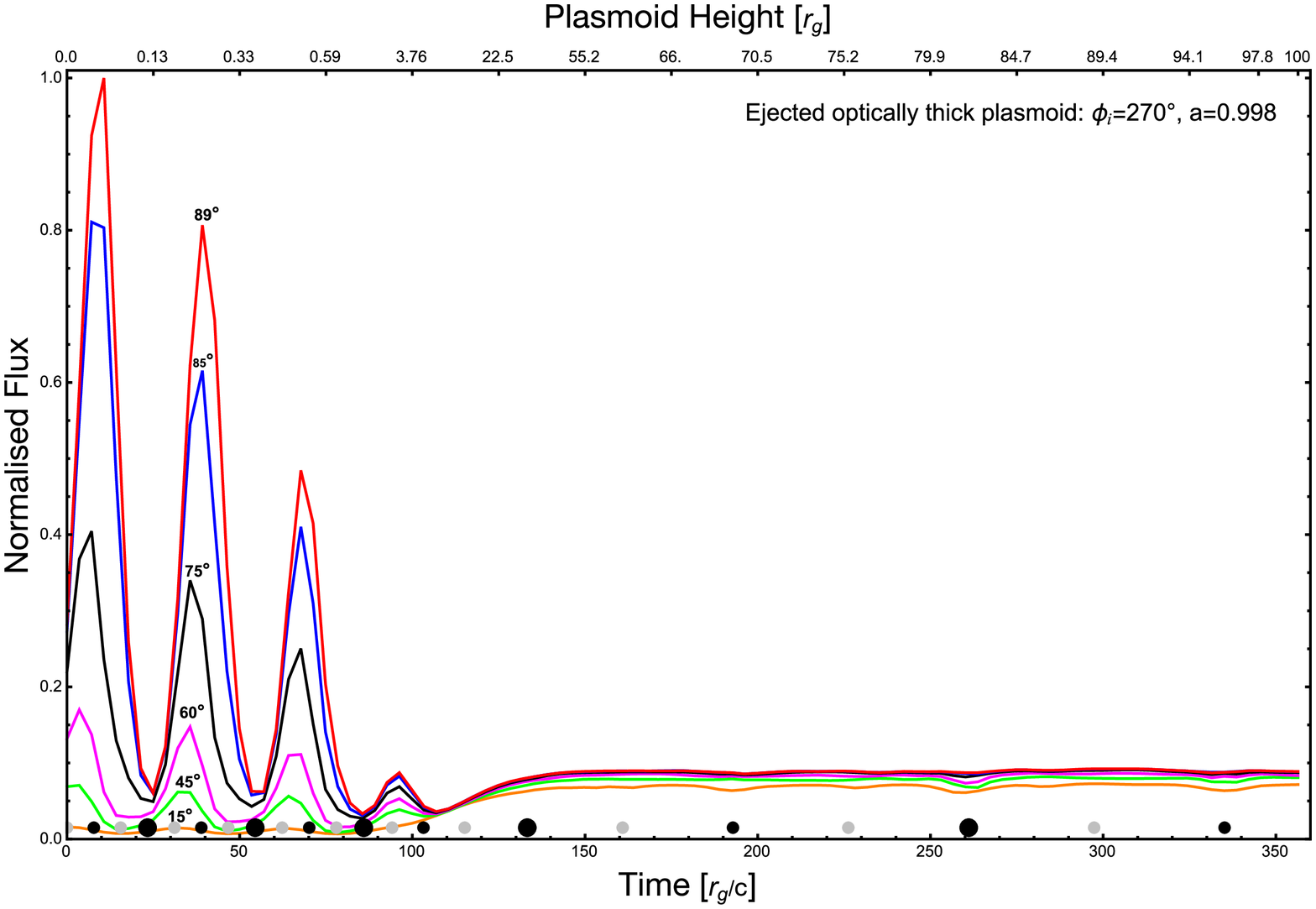}
\caption{As in Figure \ref{fig-12} but now considering a Kerr ($a=0.998$) black hole.} 
\label{fig-13}
\end{center}
\end{figure}

In Figure \ref{fig-13} the plasmoid now orbits and is subsequently ejected from within the vicinity of the event horizon of a Kerr black hole.
Due to the much shorter orbital timescale, the plasmoid performs more than three complete orbits 
   before it enters the the rapid acceleration phase. 
After this phase the lightcurves also flatten and become almost featureless, 
   as shown in Figure \ref{fig-12}. 
However, the amplitude and morphology of the lightcurves in both panels of Figure \ref{fig-13} are very similar 
  and therefore weakly dependent on the launching location of the plasmoid. 
Thus, the dynamical timescale of the plasmoid is more dominant in dictating  
  the temporal behaviours of the plasmoid emission as seen by the observer. 
Differences in the emission from the plasmoid at various azimuthal positions 
  do not have time to manifest and so the emission is in a sense smeared over the entire plasmoid motion.

Unlike in the case of orbiting optically thick plasmoids, 
  the viewing inclination of the system cannot be reliably inferred from 
  the the emission lightcurve of an ejected optically thick plasmoid. 
However, a constraint on the spin of the black hole can be obtained  
   from the timescale over which the peak flux of the lightcurve decays 
   prior to rapid acceleration phase of the ejected plasmoid, where the lightcurve flattens, 
   since close to the ISCO the plasmoid's orbital dynamics are determined by the black hole's rotation.  
  
\subsection{Ejected optically thin plasmoids}
 
\begin{figure}
\begin{center} 
  \includegraphics[width=0.47\textwidth]{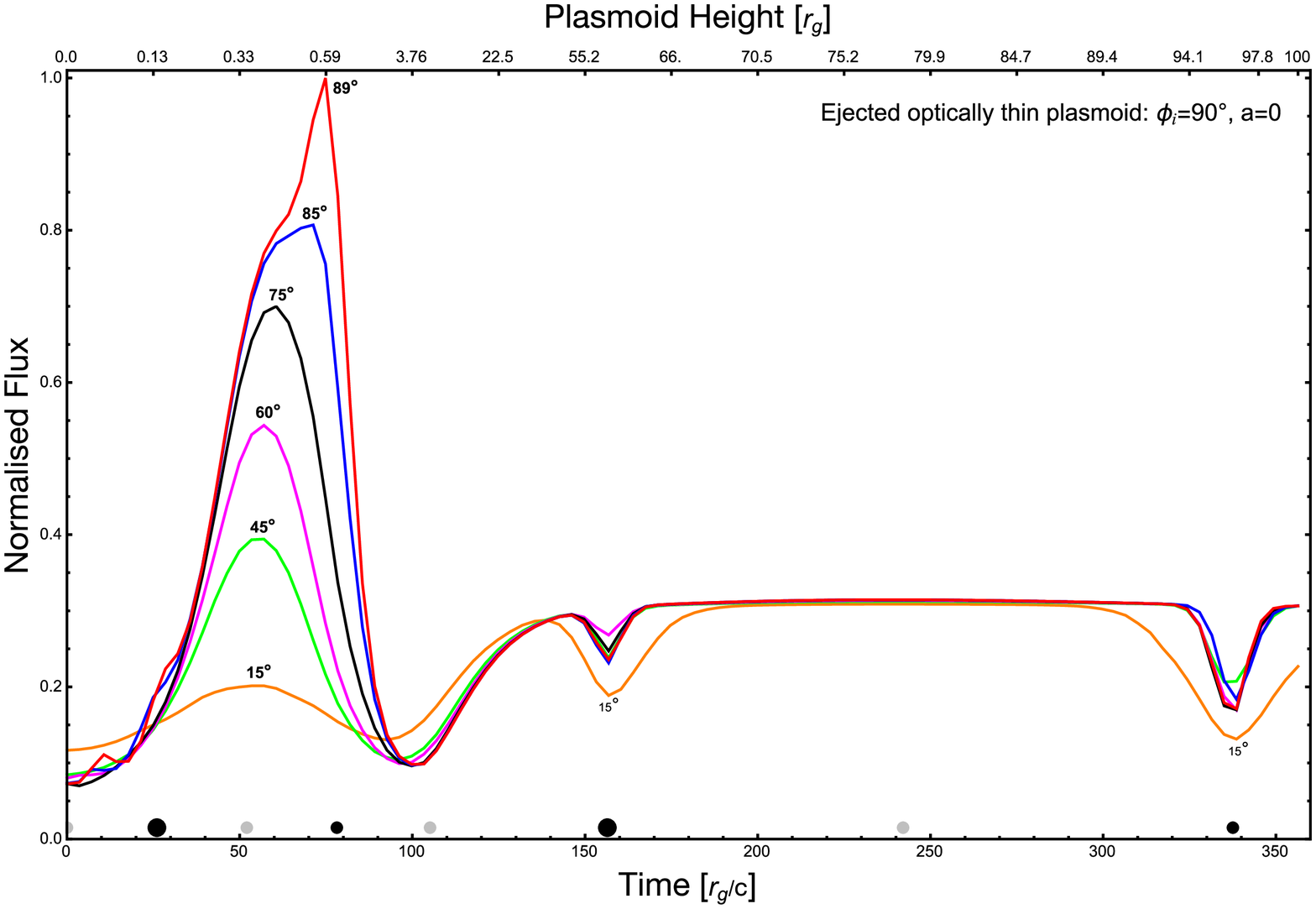}\vspace{1mm}
  \vspace*{0mm}
  \includegraphics[width=0.47\textwidth]{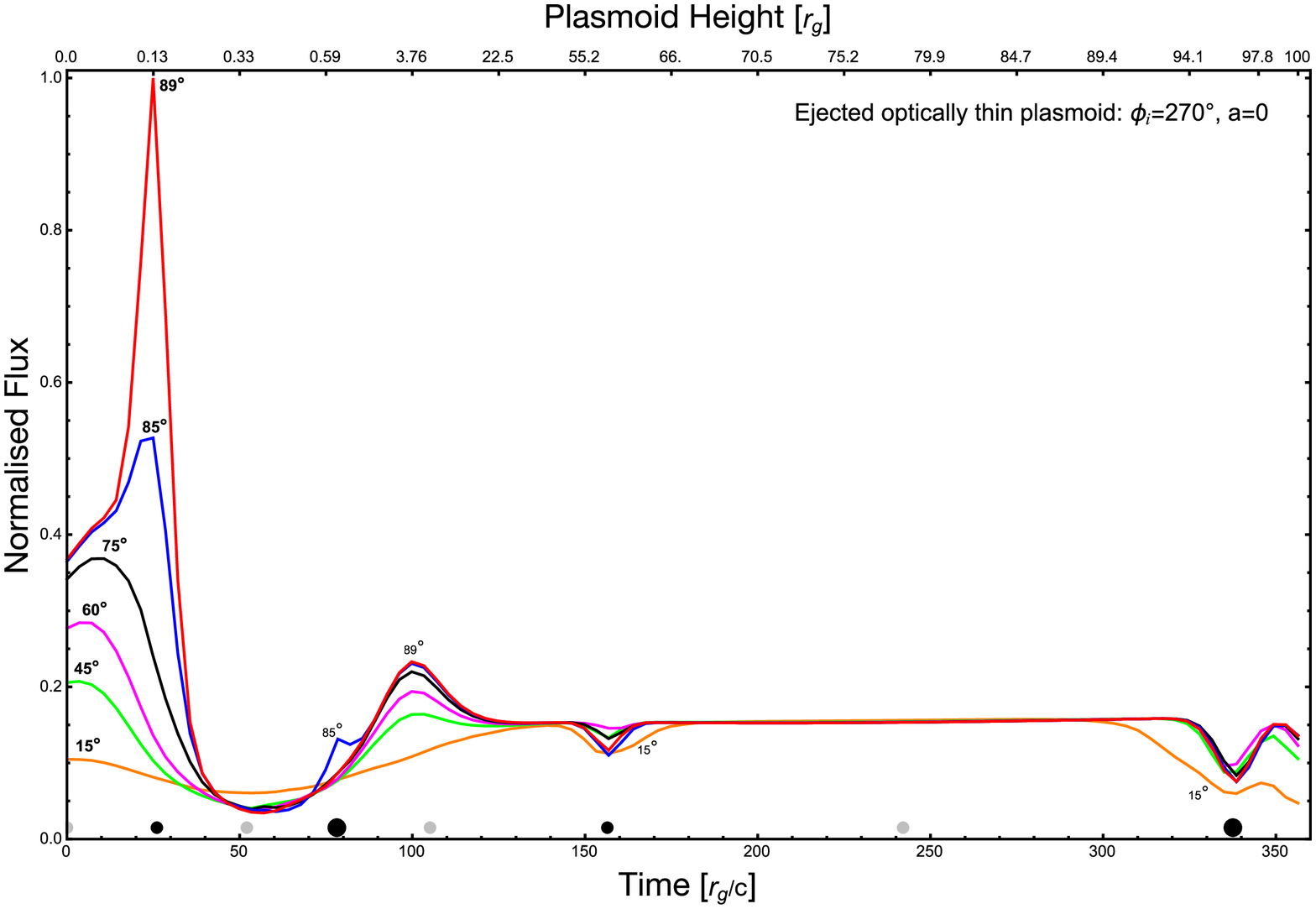}
\caption{Frequency-integrated lightcurves from an optically thin plasmoid ejected from a Schwarzschild black hole starting from an initial position of $\phi_{i}=90^{\circ}$ (top panel) 
and an initial position of $\phi_{i}=270^{\circ}$ (bottom panel). All image orders are considered.} 
\label{fig-14}
\end{center}
\end{figure}

\begin{figure}
\begin{center} 
  \includegraphics[width=0.47\textwidth]{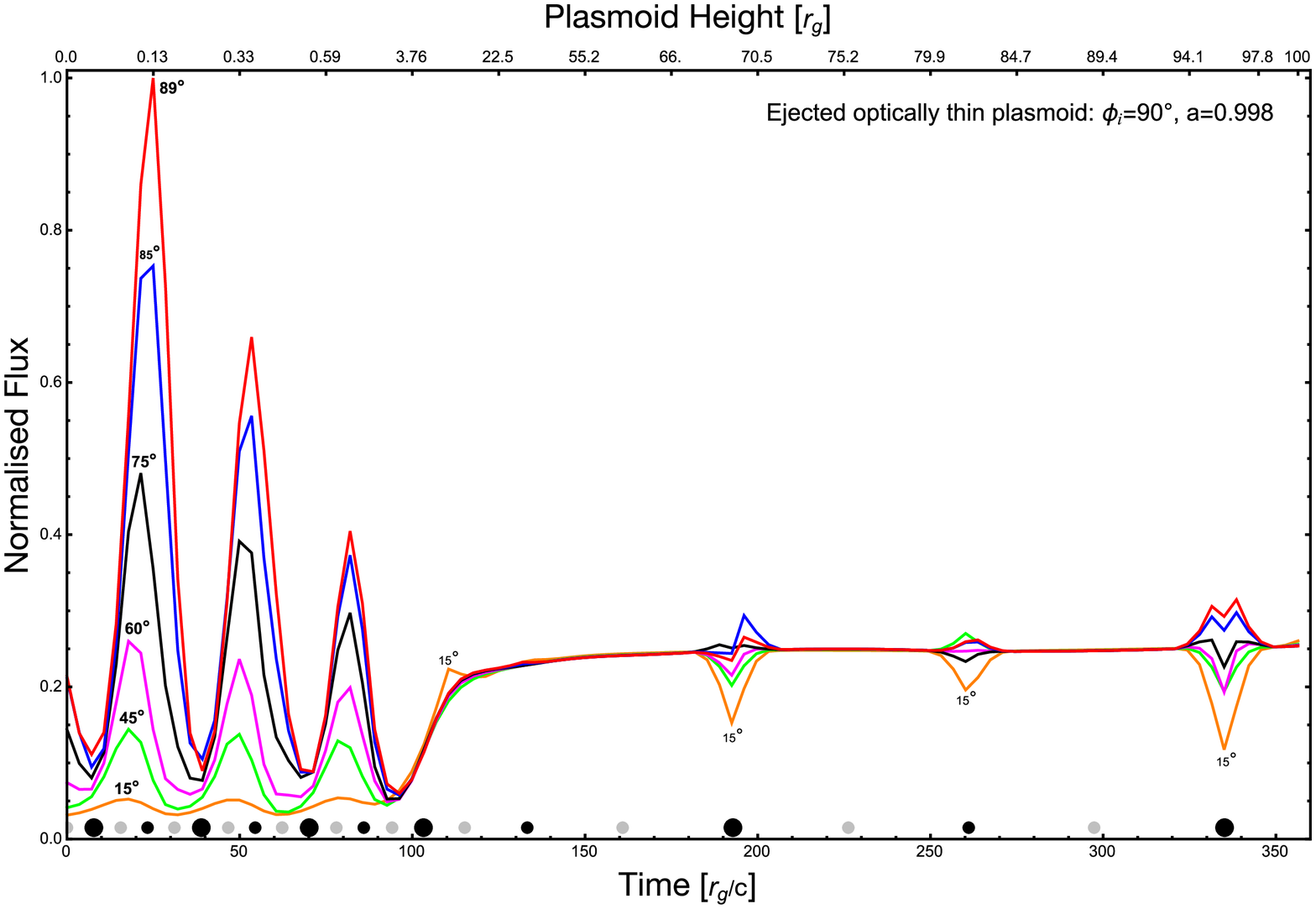}\vspace{1mm}
  \vspace*{0mm}
  \includegraphics[width=0.47\textwidth]{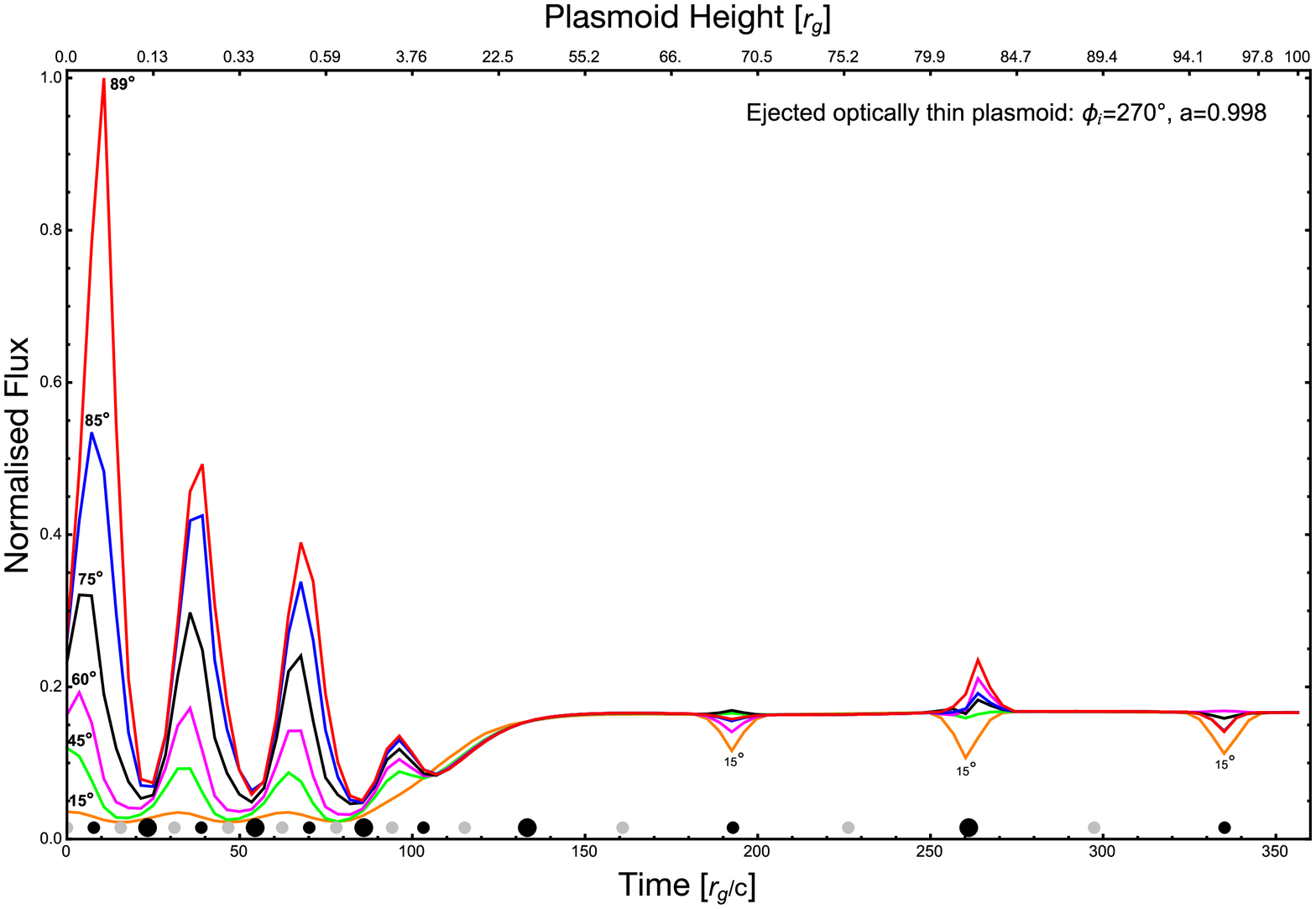}
\caption{As in Figure \ref{fig-14} but now considering a Kerr ($a=0.998$) black hole. } 
\label{fig-15}
\end{center}
\end{figure}

As seen in \S 3.2, 
  comparisons of lightcurves of optically thin and optically thick orbiting plasmoids 
  have shown that lightcurves which were very similar 
  as differences in the emission models 
  were ``smeared" over the entire orbital phase, 
  yielding a profile akin to an average of the emission over one orbital period. 
This is expected to be and is indeed the case for an ejected optically thin plasmoid 
  prior to the rapid acceleration phase, when its height changes rapidly.
It can be seen in all panels of Figures \ref{fig-14} and \ref{fig-15}  
   (for a Schwarzschild and a Kerr black hole with $a = 0.998$ respectively) 
   that optically thin ejected plasmoids exhibit significant peaks and troughs in their lightcurves, 
   when the plasmoid moves in front of, and behind, the black hole 
   (i.e.\ directly along the observer line of sight, in front of or behind the black hole).
For the optically thin plasmoids, 
   these transverse Doppler effects are much more pronounced, 
   owing to the fact that the entire plasmoid contribute to the emission 
   that is observed by the observer.

The morphology of the lightcurves of the ejected plasmoids 
  depend on the location of the launch location 
  not only in the initial phase but also after the launch.  
Significant differences between the relative amplitude of the emission 
  (when the flux is normalised to the peak amplitude) in the post-ejection phase 
  can be seen in Figure \ref{fig-14} for the two different plasmoid positions. 
The difference in the relative amplitudes can also be seen 
   in the ejected optically thick plasmoid in the case of the Schwarzschild black hole 
   (cf.\ Figure \ref{fig-12}). 
However, for rapidly rotating Kerr black holes 
  the emission from the optically thin ejected plasmoid 
  is markedly different to the emission from its optically thick counterpart, 
  (cf.\ the lightcuves in Figures \ref{fig-13} and \ref{fig-15}).
The relative amplitude of emission of the plasmoids lauched at different locations  
  differ by $\sim25\%$ in the post-ejection phase, 
  as shown in the lightcurves in the two panels of Figure \ref{fig-15}. 
  This is a result of the entire volume of the ejected plasmoid contributing to the emission. 
Thus, the relative amplitudes of the peaks, dips and flattened parts of lightcurves 
  could provide clues as to whether the plasmoid is optical thick or optically thin 
  in the corresponding emission processes 
  (as measured in the local reference frame of the plasmoid).  

\section{Discussion and Conclusion}  

\subsection{Summary of the results}  

The formulation for the radiative transfer calculations in this work is fully covariant. 
It accounts for various gravitational effects due to the black hole 
   as well as the relativistic effects due to the motion of the plasmoid. 
In the construction of the emission lightcurves, 
  specific time-keeping and correction procedures have been employed 
  to extract photons from the plasmoid images generated by the ray-tracing and radiative transfer calculations. 
These procedures ensure that the arrival time of each of the photons 
  in the beamed and (multiple-order) lensed emission 
  from the (same) plasmoid at different locations is registered correctly.  

Here we summarise the main results of our calculations. \\ 
(i) The plasmoid's emission lightcurves exhibit complex temporal behaviour as seen by an observer, 
       despite the fact that the emission is constant with respect to time in the plasmoid's local reference frame.  
     The brightness variations are 
        due to gravitational effects of the black hole (in particular, gravitational lensing)
        which combine with special-relativistic effects of the plasmoid's motion,  
        (such as beaming and Doppler boosting),
        as well as the convolution of these two kinds of effects with the viewing geometry of the system.   \\ 
(ii) Gravitational lensing modifies the optical path lengths of the emission from the plasmoid to the observer,   
    and hence the emission from different parts of the plasmoids have differential arrival times. 
   At any instant (in the reference frame of the observer),  
     an image  
     would consist of lensed emission from the plasmoid with many different lensing orders.  
  To account for the effects from the differential arrival times of the photons  
    and the superposition of lensed emission of multiple image orders, 
    proper time-keeping and time correction procedures are needed  
    when extracting the photons from the images to construct the emission lightcurve of the plasmoid.     \\  
(iii) Relativistic beaming and Doppler boosting due to the motion of the plasmoid 
      cause amplification or suppression of the emission.  
     This introduces brightness variations in the (bolometric) lightcurves of the plasmoids. 
    Additional complications in the brightness variations will result  
      if the lightcurve is constructed within a finite photon energy/frequency bandwidth, 
      because of the association of differential arrival times with energy shifts of photons in 
      the emission, when they are evaluated in the reference frame of the observer.  \\ 
(iv) In the case of the Kerr black hole, 
        rotational frame dragging can give rise to a varienty of lightcurve morphologies.  
    It can smear an emission beam that is caused by the plasmoid's relativistic motion.   
    Beam smearing will reduce the amplitudes of brightness variations in the lightcurve.  
    It can also convolve with gravitational lensing, 
        leading to differential energy/frequency shifts of the photons. \\   
(v) Gravitational lensing and rotational frame-dragging (due to the black hole's gravity),  
      can at times compete rather than convolve with 
      relativistic beaming and Doppler boosting (due to the motion of the plasmoid).  
   For the ejected plasmoids, relativistic effects eventually dominate.
        Since the gravitational field strength diminishes as the plasmoid is increasingly further from the black hole. 
      special-relativistic effects arising from the plasmoid's motion become more prevalent. \\ 
(vi) The morphology of the plasmoid lightcurves are dependent on the viewing inclination angle of the system, 
       as gravitational lensing and relativistic boosting 
       are generally more prominent at higher viewing inclination angles. 
   Viewing inclination, however, becomes less relevant for variations in the emission from the ejected plasmoids  
      for distances sufficiently far from the black hole event horizon. \\              
(vii) For the optically thick plasmoids, 
        the emission is from the fraction of the plasmoid surface ``visible'' to the observer. 
   For the optically thin plasmoids, 
        the entire plasmoid contributes to the emission that reaches the observer. 
    As gravitational and relativistic effects convolve,
      not only the viewing geometry of the system 
      but also the optical depth of the plasmoid modify the emission.
      The lightcurves of the optically thick and the optically thin plasmoids 
      do not generally resemble each other.
            
\subsection{Remarks and astrophysical implications}  

\subsubsection{Emissivity of the plasmoids}

To investigate how gravitational and special-relativistic effects 
    introduce brightness variations in the plasmoid's observed lightcurve 
    we have adopted a simple prescription 
    within which the plasmoid's emission is isotropic in the local reference frame. 
Anisotropy, such as beaming, only arises because of the relativistic motion of the emitter 
   with respect to the observer. 
Moreover, 
   the timescales on which the energy content and the populations of emitters evolve 
   are assumed to be longer than the dynamical timescales of the system 
   and the time window within which the lightcurve is constructed. 
Thus, the emission from the plasmoid throughout an orbit or the launching process  
  can be treated as constant with respect to time. 
   
The emission from an optically thin, magnetically confined hot plasma is anisotropic, 
  as synchrotron radiation (for the radio emission) depends on the magnetic field configuration, 
  and unsaturated Compton scattering (for the X-rays) depends 
  on the microscopic and the bulk momentum distributions of the energetic electrons.  
These anisotropic emissions will be subject to different degrees of light aberration and lensing 
  when the plasmoid is at different azimuthal locations with respect to the observer. 
Also, the emissivity of such a plasmoid is not constant. 
The energy distribution and number of energetic electrons within the plasmoid evolve 
  because of particle injection, diffusion losses, acceleration and re-acceleration, 
  radiative and adiabatic cooling, and energy exchange between the electrons and other particle species. 
Developing a self-consistent model for the time-dependent evolution of emission from the plasmoid   
  would require solving the radiative transfer equation 
  together with the energy transport equations for different species of charged particles,  
  along with the structural evolution and dynamical evolution equations of the plasmoid. 
Although such calculations will providing more reliable models 
  to explain the variable emission observed in compact sources e.g..\ the flares from Sgr A* at different wavebands, 
  these complexities are beyond the scope of this work and warrant an independent study. 

In spite of this, the results, e.g.\ the lightcurves, shown in \S3 and \S4,  
   still provide very useful insights into interpreting flares and variations in the emission 
   from plasmoid ejections that evolve.    
Here is an illustration.
Flares associated with solar CMEs \citep[see][]{Aschwanden2009} 
  often have a fast-rise slow-decay profile \citep{Li2005}, 
  and such brightness profiles are also found in flares in the accretion disk around a protostar or a T Tauri star 
  \citep[see][]{Hayashi1996,Uzawa2011}.
If the flares associated with a solar CME-like plasmoid ejection 
  in an accreting disk around a black hole also have intrinsic fast-rise slow-decay profiles, 
  what signatures would we expect to see in the lightcurve?  
For a Newtonian system, a time-dependent emission 
  could retain its intrinsic characteristic temporal profile in the observed lightcurve,  
  provided that instrumental distortion is not severe, 
  i.e.\ the signature of fast-rise slow-decay would be recovered in the observational data. 
This, however, will not be guaranteed for flares associated with a plasmoid ejection 
  that occurs in the vicinity of a black hole.  
If the timescales of the brightness variations of a flare are
  comparable with the dynamical timescales of the system,  
  the stretching and compression of the time variabilities 
  will be very sensitive to the location of where the variabilities arise. 
In this situation, 
  gravitational lensing and time-dilation can stretch the initial rise phase 
  while relativistic beaming can compress the subsequent decay phase,     
  when the plasmoid, which is launched at a certain azimuthal location at the inner accretion disk,  
  is viewed at a high inclination angle.  
A fast-rise slow-decay flare could lose its characteristics,  
   and it may even appear to be time-reversal symmetric, 
  or as slow-rise fast-decay, in the observations. 
 
\subsubsection{Dynamics of plasmoid ejection} 

A plasmoid is an extended body. 
In our calculations the plasmoid is assumed to be rigidly held together,
 with all material within moving at the same velocity. 
How does this assumption affect the results that we have obtained? 
Studies \citep[see][]{Schnittman2004, Broderick2005, Schnittman2006, Schnittman2006MHD, Yang2013, Li2014},  
   indicate that the spectrograms and lightcurves from an emitting plasma blob   
   in the vicinity of a black hole 
   are weakly dependent on the size of the emitter.    
However, a finite-size plasmoid near a black hole will subject to a strong tidal force, 
   which will cause it to deform over the course of its orbital motion around the black hole.     
This tidal effect is dynamically important 
   \citep[see][]{Mashhoon1975, Carter1983,Christian2015}.
There is much scope to improve the underlying dynamical model of the plasmoid itself. 
The model adopted in this study \citep[see][]{Yuan2009} 
   is essentially Newtonian, 
   and it only accounts for the vertical motion of the plasmoid.    
Tracking and resolving magnetic reconnection events and their subsequent flaring 
  within GRMHD simulations \citep[e.g.][]{DeVilliers2003, Moscibrodzka2014, Chan2015} 
  will naturally provide a more realistic dynamical model 
  for the plasmoid motion and ejection.
Nevertheless, self-consistently coupling such simulations with radiative transfer calculations
  is beyond the scope of this work 
  and it is more appropriate to leave it for future studies.

\subsubsection{Further remarks}

So far we have considered isolated plasmoids before or after launching 
  and have ignored the ``background'' emission from the other components in the system 
  when constructing the lightcurves.    
The emission from the accretion disk itself may vary. 
There are also emissions from structures associated with (but external to) the plasmoid, 
  such as the protruding magnetic arcade hot spots in the accretion disk. 
The formulation for the general-relativistic radiative transfer calculations is generic.  
The results that we have obtained from the radiative transfer calculations 
  for the emission from orbiting and ejected plasmoids 
  can easily be generalised and 
  provide insights into broader astrophysics problems, 
  in particular those concerning time variabilities in emission 
  from structures in relativistic accretion disks and outflows near a black hole. 

Analogous to the solar corona, accretion disk coronae exhibit a variety of eruptive magnetic activities 
   \citep{Beloborodov1999}. 
In the CME model for plasmoid ejections modeled in this work, 
  before the full development of the plasmoid and its subsequent ejection, 
  magnetic filaments and flux ropes must have been formed in the accretion disk corona 
   \citep[cf.\ solar CME formation, see][]{Chen2011,Janvier2013}. 
Similar to those in the solar corona, 
  these magnetic arcades confine hot thermal plasmas and generate non-thermal energetic charged particles,   
  through shocks within, and plasma and magnetohydrodynamics processes in the interfacing region, 
  where the footpoints of the aracades anchor onto the accretion disk.  
These magentic features emit radiation as plasmoids.  
If the linear extent of the magnetic arcades are small 
  (relative to the length scale on which the local gravitational curvature radius os defined), 
  they may be treated as a (spherical) plasmoid as a first approximation. 
Since they are optically thin to the radiative processes, 
   the temporal properties of their emission 
   would resemble those of the orbiting optically thin plasmsoid (see \S3.2), 
   although in this case the anisotropy of the emission 
   would need to be considered in the radiative transfer calculations.  
 
 Hot spots in the accretion disk 
  can be generated by streaming particles and high-energy radiation  
  from the plasmoid and the magnetic arcades, 
  or by irradiation of the X-rays from the plasmoid.  
Magnetic reconnection associated with CMEs can produce extremely energetic charged particles.  
 They stream down following the magnetic field lines in the flux ropes
  and deposit their energies into the denser surface layer of the accretion disk,
  creating (two) hot spots at the flux-rope footpoints   
   \citep[cf.\ solar CME, see][]{Brown2009,Lin2011,Gordovskyy2014}. 
Thermal free-free X-rays will be emitted from these hot spots,   
  and the emission will be gravitationally lensed and relativistically boosted. 
 Moreover, each hot spot is a 2D surface.  
Thus, the emission from the hot spots will exhibit temporal properties 
  similar to the emission from orbiting (opaque) plasmoids. 
The formation processes of hot spots by irradiation of hard X-rays from the plasmoid 
  are similar to the formation of hot layers by external irradiation of accretion disks 
  in accreting compact objects, 
  except that the former is a more localised effect. 
The hot spots produced by hard X-ray irradiation of the plasmoid 
  would show spectra with edges and line features \citep{Matt1993,Magdziarz1995,Ross1999}, 
  which will be subject to gravitational and relativistic effects \citep[see e.g.][]{Matt1993,Lee2009}.  
 The emission from these irradiation induced hot spots will show temporal properties 
   similar to those of the emission from orbiting opaque plasmoids. 
   
The CME plasmoid ejection model was invoked originally to explained the episodic outflows  
   from galactic centres, e.g. Sgr A* \citep{Yuan2009}. 
Here we show explicit radiative transfer calculations of the emission from the plasmoid 
  for various system configurations. 
In reality, the CME plasmoid could co-exist with a more continuous large-scale outflow, 
  such as the GRMHD flow described in \cite{Pu2015}. 
The emission from the plasmoid would superimpose 
  with the emission from the large-scale outflow, 
  which will be observed to show intrinsic variabilities 
  due to instabilities within the flow or wave formation in the flow. 
If the emission of the large-scale flow fluctuates or varies 
  on timescales similar to the dynamical timescale of the plasmoid, 
  then it will become non-trivial to entangle 
  the variabilities of the emission from the plasmoid ejections and from the large-scale flow. 

\subsection{Conclusion}

Gravitational and relativistic effects on emission from orbiting and ejected plasmoids
  in the vicinity of a black hole waere investigated. 
These effects introduced time variability in the plasmoid's observed emission. 
Lightcurves of the plasmoid's emission 
  at various plasmoid locations and viewing geometries of the system  
  were calculated 
  for a Schwarzschild and a Kerr back hole.   

For the plasmoid in orbit close to the black hole, 
  the emission is smeared over the sky due to gravitational lensing.
At high inclination angles, gravitational lensing can produce an Einstein ring, 
  magnifying the emission from the plasmoid by factors of $\sim10-100$. 
In the corresponding frequency-integrated lightcurves, as inclination angle increases, the emission 
  becomes sharply peaked due to relativistic beaming and gravitational lensing. 
This also causes the lightcurves to be skewed towards later times. 
At the highest inclinations, gravitational lensing dominates over relativistic beaming,  
  and the lightcurves becomes double-peaked. 
The lightcurves are very sensitive to the orbital period of the plasmoid, 
  and hence the spin of the black hole, as well as observer inclination angle. 
The presence of separated double peaks in the observed lightcurves 
  can potentially constrain the black hole spin.

A dynamical and kinematic model based on a solar CME-like magnetically-driven plasmoid ejection 
  was then employed to determine motion of the plasmoid as the ejection progressed.  
With the plasmoid position and momentum at each point in time specified, 
   the emission from the plasmoid was calculated, 
 and the observed lightcurves 
  during the first $100\, r_{\mathrm{g}}$ of its vertical motion were constructed. 
We have found that the lightcurves were sensitive to initial azimuthal position, 
  either greatly enhancing or suppressing the appearance of the rapid acceleration phase of the plasmoid. 
The rapid acceleration phase timescale was found to be consistent across all lightcurves 
  and independent of intrinsic black hole parameters. 
Near the event horizon, 
  when the orbital timescale is of the order of the light crossing timescale of the system, 
  photon time delay effects became significant.
By properly and self-consistently accounting for photon arrival time effects  
  convolved with all relativistic effects 
  we obtained lightcurves distinct from conventional uncorrected lightcurves.
Relative time delays between photons, between photons and the emitting medium, 
  and between photons and the observer are extremely important effects
  which have many consequences for the observed emission in these systems.

The general-relativistic radiative transfer calculations 
  and the procedures for lightcurve construction 
  presented in this work are generic. 
The results from the calculations of time variabilities in the observed emission 
  from orbiting and ejected plasmoids 
  provide useful insights for the study of related relativistic systems, 
  such as emission from eruptive magnetic flares 
  or hot spots in relativistic accretion disks around black holes.

\section*{Acknowledgements}
We thank P.F.~Chen and D.~Baker 
  for discussions on the theoretical and observational aspects 
  of coronal eruptive processes in the Sun.  
Z.Y.\, is supported by an Alexander von Humboldt Fellowship 
  and acknowledges support from the ERC Synergy Grant 
  ``BlackHoleCam -- Imaging the Event Horizon of Black Holes" (Grant 610058). 
This research has made use of NASA's Astrophysics Data System. 

\bibliographystyle{mn2e}
\bibliography{references.bib}

\label{lastpage}
\end{document}